\documentclass[twocolumn]{autart}    

\usepackage{graphicx}          
\usepackage{amssymb}
\usepackage{amsmath}
\usepackage{enumitem}
\usepackage{mathrsfs}
\usepackage{algorithm}
\usepackage{multirow}
\usepackage{algpseudocode}
\usepackage{booktabs}
\newtheorem{theorem}{Theorem}{}
\newtheorem{proposition}{Proposition}{}
\newtheorem{lemma}{Lemma}{}
\newtheorem{assumption}{Assumption}
\newtheorem{corollary}{Corollary}{}
\newtheorem{remark}{Remark}{}

\begin{document}

\begin{frontmatter}

\title{Deep Koopman Iterative Learning and Stability-Guaranteed Control for Unknown Nonlinear Time-Varying Systems}
\thanks[footnoteinfo]{This paper was not presented at any IFAC 
meeting. Corresponding author: Zhisheng Duan(duanzs@pku.edu.cn).}

\author[PKU]{Hengde Zhang}\ead{zhanghengde@stu.pku.edu.cn},    
\author[PKU]{Yunxiao Ren}\ead{renyx@pku.edu.cn},    
\author[PKU]{Zhisheng Duan}\ead{duanzs@pku.edu.cn},
\author[PKU]{Zhiyong Sun}\ead{zhiyong.sun@pku.edu.cn},
\author[CityU]{Guanrong Chen}\ead{eegchen@cityu.edu.hk}

\address[PKU]{School of Advanced Manufacturing and Robotics, Peking University, Beijing, China}  
\address[CityU]{Department of Electrical Engineering, City University of Hong Kong, Hong Kong, China}             
          
\begin{keyword}                           
Deep neural network; Koopman operator; Nonlinear time-varying system; System identification; Data-driven control.               
\end{keyword}                             

\begin{abstract}                          
This paper proposes a Koopman-based framework for modeling, prediction, and control of unknown nonlinear time-varying systems.
We present a novel Koopman-based learning method for predicting the state of unknown nonlinear time-varying systems, upon which a robust controller is designed to ensure that the resulting closed-loop system is input-to-state stable with respect to the Koopman approximation error.
The error of the lifted system model learned through the Koopman-based method increases over time due to the time-varying nature of the nonlinear time-varying system.
To address this issue, an online iterative update scheme is incorporated into the learning process to update the lifted system model, aligning it more precisely with the time-varying nonlinear system by integrating the updated data and discarding the outdated data.
A necessary condition for the feasibility of the proposed iterative learning method is derived.
In order to reduce unnecessary system updates while ensuring the prediction accuracy of the lifted system, the update mechanism is enhanced to determine whether to update the lifted system and meanwhile to reduce updates that deteriorate the fitting performance.
Furthermore, based on the online-updated lifted system, a controller is designed to ensure the closed-loop controlled system be input-to-state stable with respect to the Koopman approximation error.
Numerical simulations on the Duffing oscillator, the serial manipulator, and the synthetic biological network system are presented to demonstrate the effectiveness of the proposed method for the approximation and control of unknown nonlinear time-varying systems.
The results show that the proposed approach outperforms existing methods in terms of approximation accuracy and computational efficiency, even under significant system variations.
\end{abstract}

\end{frontmatter}

\section{Introduction}
In recent years, data-driven control has emerged as a fast-evolving direction of research within the field of systems control, demonstrating clear advantages over traditional control methods.
As real-world control systems become increasingly complex and modeling becomes more challenging, the data-driven approach, which does not require precise system models, is becoming more promising and practical.
In particular, Koopman operator theory \cite{koopmanHamiltonianSystemsTransformation1931} is well-known for its ability to apply effective techniques from linear systems to address nonlinear systems and control \cite{koopmanHamiltonianSystemsTransformation1931,huangFeedbackStabilizationUsing2018,kordaLinearPredictorsNonlinear2018,zhangRobustTubebasedModel2022,kordaOptimalConstructionKoopman2020,mauroyKoopmanOperatorSystems2020}.
Koopman operator theory describes a nonlinear system using an infinite-dimensional linear operator that acts on a vector space of functions.
In order to make the infinite-dimensional Koopman operator applicable to real-world scenarios, a number of approximation methods have been developed and studied.
Analogous to identifying linear systems from data using Dynamic Mode Decomposition (DMD) \cite{kutz2016dynamic,zhangOnlineDynamicMode2019}, Extended DMD (EDMD) \cite{williamsDataDrivenApproximation2015,kordaConvergenceExtendedDynamic2018} has been proposed to identify finite-dimensional Koopman operators.
EDMD and its variants \cite{haseliGeneralizingDynamicMode2023,haseliLearningKoopmanEigenfunctions2022,shiACDEDMDAnalyticalConstruction2022} provide effective methods to approximate the Koopman operator on a predefined finite-dimensional space spanned by a dictionary of functions, often called observables.
Subsequently, extensive researches \cite{kordaLinearPredictorsNonlinear2018,zhangRobustTubebasedModel2022,shiDeepKoopmanOperator2022,jiaEVOLVEROnlineLearning2024} have been conducted on the data-driven identification and control of nonlinear systems based on EDMD.

Recent works \cite{haoDeepKoopmanLearning2024,zhangOnlineDynamicMode2019,chamorro2021line,gueho2021time} have focused on dealing with time-varying nonlinear systems.
In \cite{haoDeepKoopmanLearning2024}, a deep Koopman learning method is developed for predicting the states of time-varying nonlinear systems and analyzing the error bound of the system state estimation. 
That work provides an iterative update version of the standard EDMD.
However, it exclusively focuses on continuously updating the higher-dimensional system with new data points, without considering the removal of outdated data points.
In some studies \cite{zhangOnlineDynamicMode2019,chamorro2021line}, simulations or experiments demonstrate that the DMD method with a forgetting factor or sliding window is more suitable for identification of time-varying systems.
In \cite{zhangOnlineDynamicMode2019}, DMD and its weighted variants are employed for system identification of time-varying linear systems. 
It demonstrates that the data-driven system identified using DMD with a forgetting factor or sliding window yields eigenvalue variations that more closely match those of the original linear time-varying system, compared to the standard DMD algorithm.
The work in \cite{chamorro2021line} employs a sliding-window Koopman Mode Decomposition (KMD) approach for online coherency analysis of power systems.
However, the above works \cite{zhangOnlineDynamicMode2019,chamorro2021line} overlook the necessary conditions that the sliding window should satisfy.
Notably, the above studies focus on the online identification of time-varying systems to obtain an updated surrogate model, without further utilizing the model for control applications.

To address the above-mentioned issues, this paper presents a theoretical framework for online iterative learning and control of unknown nonlinear time-varying systems utilizing deep Koopman operators. 
The contributions of this paper are summarized as follows:
\begin{itemize}
    \item First, we propose a novel online iterative update method for the lifted system representation. This method dynamically integrates new data while discarding outdated data, enabling the model to adapt to time-varying system dynamics. Comparative experimental results demonstrate that our approach achieves a significant reduction in prediction error compared to existing state-of-the-art methods \cite{haoDeepKoopmanLearning2024}.
    \item Second, we establish a necessary condition for the iterative feasibility of updating the lifted system matrices and provide an upper-bound estimation for the resulting estimation error. Furthermore, we introduce two innovative update mechanisms to govern the system update process. These mechanisms not only drastically reduce the computational complexity by reducing the number of updates but also, as verified by experimental results, further enhance the prediction accuracy.
    \item Third, we design a robust controller based on the iteratively updated lifted system. We rigorously prove that the resulting closed-loop system, which incorporates the original physical plant, is input-to-state stable with respect to the Koopman approximation error, thus guaranteeing its stability in applications.    
\end{itemize}
The method trains a neural network using the input-output data from the early stage of the nonlinear time-varying system's operation.
Thereafter, the lifted system is updated online in an iterative manner to achieve accurate state prediction.
Based on this online-updated model, a controller with stability guarantee is designed to perform tracking tasks with high precision.

Compared with the deep Koopman representation method in~\cite{haoDeepKoopmanLearning2024}, our approach explicitly discards outdated data during the iterative update process.
Compared with the online DMD method in~\cite{zhangOnlineDynamicMode2019}, we provide necessary conditions under which the iterative updates can be carried out.
Compared with the time-varying eigensystem realization approach in~\cite{gueho2021time}, our method updates the Koopman operator matrices using online-collected data, enabling adaptation to system variations.
Simulation results\footnote{Code available at: {https://github.com/wulidede/Online-time-varying-deep-Koopman-learning.git}.} further demonstrate that the proposed approach achieves substantial reductions in both prediction error and computational cost compared with existing methods \cite{zhangOnlineDynamicMode2019,haoDeepKoopmanLearning2024}.
Moreover, while prior methods~\cite{zhangOnlineDynamicMode2019,haoDeepKoopmanLearning2024,gueho2021time} focus exclusively on state prediction, our work goes beyond prediction by developing a model predictive control method with stability guarantees based on the iteratively updated model, thereby forming a unified framework for modeling, prediction, and control of unknown time-varying nonlinear systems.

The organization of this paper is as follows: Section~\ref{sec:problem} presents the problem statement and the system modeling.
Section~\ref{sec:main_results} introduces the online iterative update method and establishes some working conditions.
Section~\ref{sec:controller} describes the design of a stability-guarantee controller based on the iteratively updated lifted system.
Section~\ref{sec:simulation} shows three numerical simulation examples to demonstrate the effectiveness of the proposed method.
Section~\ref{sec:conclusion} concludes the paper.

\textit{Notation.}
Let $\mathbb{R}_{\geq 0}$ and $\mathbb{Z}_{\geq 0}$ denote the set of non-negative real numbers and of non-negative integers, respectively.
For a matrix $A \in \mathbb{R}^{n \times m}$, $A^{\dagger}$ denotes its Moore-Penrose pseudoinverse. 
The notation $\|\cdot\|_W$ denotes the weighted norm, where $W$ is a positive definite matrix.
A continuous function $\alpha:[0,\infty)\to [0,\infty)$ belongs to class $\mathscr{K}$, denoted by $\alpha \in \mathscr{K}$, if it is strictly increasing with $\alpha(0)=0$. It belongs to class $\mathscr{K}_\infty$ if additionally $a=\infty$ and $\alpha(r)\to\infty$ as $r\to\infty$.
The time complexity of an algorithm is given by $\mathcal{O}(\cdot)$.

\section{Problem formulation}\label{sec:problem}
Consider a discrete-time time-varying nonlinear system with unknown dynamics,
\begin{equation}\label{eq:system}
x_{k+1} = f(x_k, u_k, k),
\end{equation}
where $x_k \in \mathbb{R}^n$ and $u_k \in \mathbb{R}^m$ denote the system state and control input at time index $k$, respectively. 
{The function $f : \mathbb{R}^n \times \mathbb{R}^m \times \mathbb{Z}_{\ge 0} \to \mathbb{R}^n$ represents the unknown nonlinear dynamics, and $k \in \mathbb{Z}_{\ge 0}$ is the discrete-time index.}

{We begin by recalling the Koopman operator framework~\cite{koopmanHamiltonianSystemsTransformation1931,kordaLinearPredictorsNonlinear2018} for analyzing an uncontrolled dynamical system of the form \(x_{k+1} = f(x_k,k)\).}
{Let $\mathcal{F} = L^2(\Omega)$ denote the $L^2$ space of real-valued functions on a compact domain $\Omega \subseteq  \mathbb{R}^n $.}
The Koopman operator $\mathcal{K}^{(k)}:\mathcal{F}\rightarrow\mathcal{F}$ is defined by
\begin{equation}
(\mathcal{K}^{(k)}\psi)(x_k) = \psi\circ f(x_k,k)= \psi(f(x_k,k)),
\end{equation}
{where $\psi\in\mathcal{F}$, $k$ is the discrete-time index and $\circ$ denotes the composition operator, i.e., $\psi \circ f(\cdot) = \psi(f(\cdot))$.}
The dynamics in the lifted state-space can be represented by a linear form,
\begin{equation}\label{eq:koopman}
    \psi(x_{k+1})=\mathcal{K}^{(k)}\psi(x_k).
\end{equation}
For controlled systems, by following the treatment in \cite{kordaLinearPredictorsNonlinear2018,haoDeepKoopmanLearning2024}, the control input \(u_k\) is linearly embedded into the lifted-space dynamics.
Select \( r \) observables from \( \mathcal{F} \) and stack them into a vector, denoted as \( g: \mathbb{R}^n\rightarrow\mathbb{R}^r \).
Then, the controlled nonlinear time-varying system (\ref{eq:system}) is transformed to the following state-space model:
\begin{equation}\label{eq:koopman_state}
    \begin{aligned}
        g({x}_{k+1}) &= A_{k}g({x}_k) + B_{k}u_k,\\
    \end{aligned}
\end{equation}
where \(A_{k} \text{ and } B_{k}\) are the lifted system matrices, which are time-varying and depend on the time index \(k\).
The lifted state is mapped back to the original state space through $x_{k+1} = C_k g(x_{k+1}),$ where $C_k$ is the matrix corresponding to the decoder.

Compared to commonly used basis functions, deep neural networks (DNNs) ---capable of representing complex functions---offer greater flexibility and have been applied in many studies \cite{shiDeepKoopmanOperator2022,haoDeepKoopmanLearning2024,luschDeepLearningUniversal2018,hanDeepLearningKoopman2020} related to Koopman operators.
In the present work, DNNs are employed as observables to obtain an approximate Koopman operator using EDMD.
By replacing $x_{t+1}$ with $\hat{x}_{t+1}$, the expression (\ref{eq:koopman_state}) can be rewritten as
\begin{equation}\label{eq:koopman_state_hat}
    \begin{aligned}
        g(\hat{x}_{k+1}, \theta_{\tau}) &= A_{\tau}g(\hat{x}_k, \theta_{\tau}) + B_{\tau}u_k,\\
        \hat{x}_{k+1} &= C_{\tau}g(\hat{x}_{k+1}, \theta_{\tau}),
    \end{aligned}
\end{equation}
where $g(\cdot, \theta_{\tau})$ is a DNN as observables with network parameter $\theta_{\tau} \in \mathbb{R}^q$, $A_{\tau}, B_{\tau} \text{ and } C_{\tau}$ are the lifted system matrices after the $\tau$-th update, and $\tau$ is the update index. 

{The \textbf{objectives of this paper} are to achieve the following:
\begin{enumerate}
    \item \textbf{Design an iterative update algorithm} based on the Koopman operator to approximate an unknown time-varying nonlinear system by a linear time-varying system, such that the predicted states remain as close as possible to those of the original system while reducing computational cost.
    \item \textbf{Develop a controller with stability guarantees} to accomplish stabilization and tracking tasks for the original time-varying nonlinear system.
\end{enumerate}}

\section{Online learning of nonlinear time-varying systems}\label{sec:main_results}
This section introduces the design of an iterative update algorithm, presenting a necessary condition for its feasibility.
Subsequently, the design of update conditions is discussed to address the issue of negative updates.

\subsection{Preliminaries}
\begin{figure}
    \begin{center}
    \includegraphics[height=4.5cm]{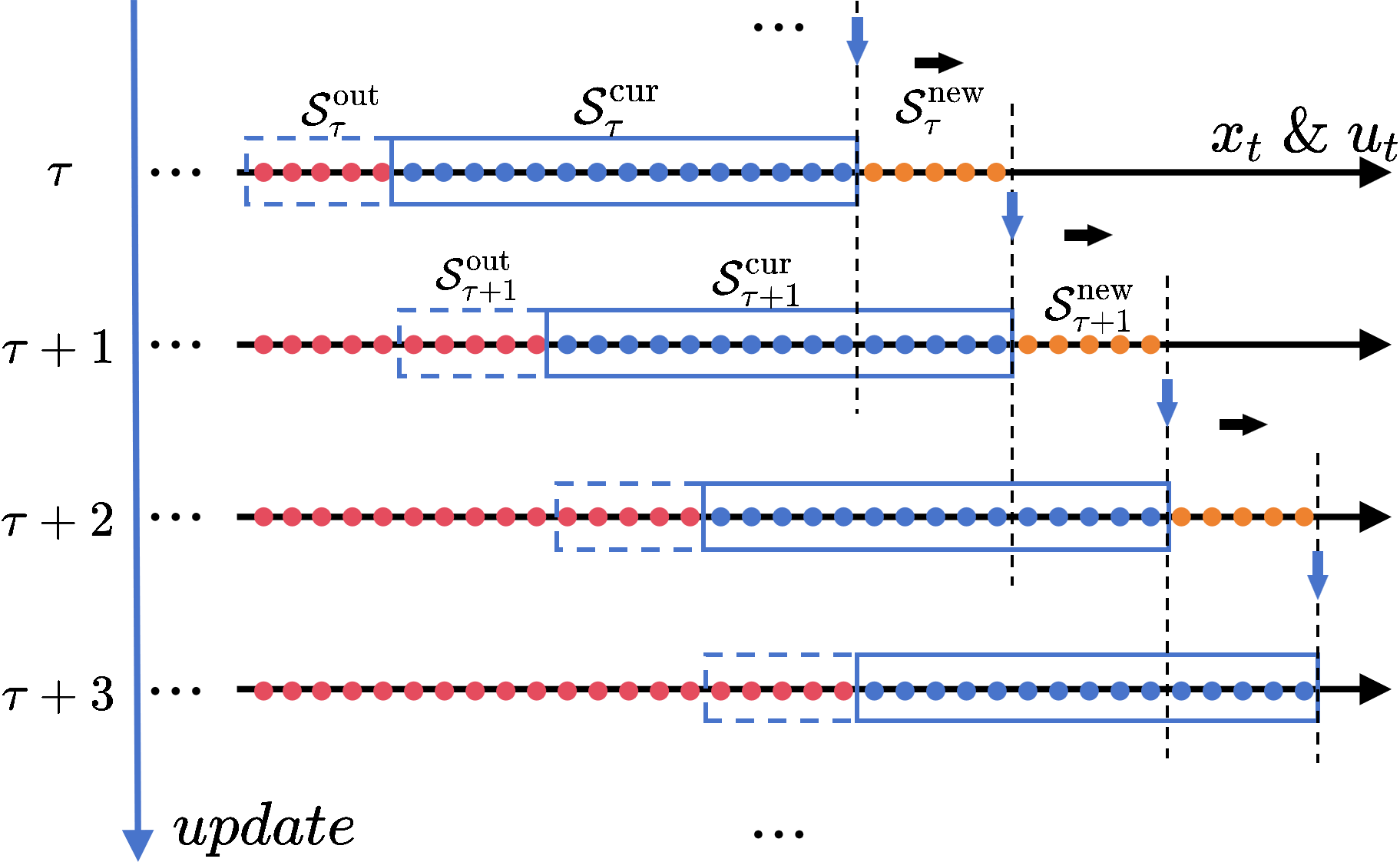}  
    \caption{Snapshots division and evolution.}  
    \label{fig:snapshots}     
    \end{center}     
\end{figure}

First, the division of input-output data (i.e., \textit{snapshots}) is clarified. 
Snapshots are sampled at equal time intervals, where \( k_\tau, k_{\tau+1}, \dots \), represent the time indices, satisfying \( k_{\tau+1} = k_\tau + b \), with \( \tau \) being the update index and $b$ a positive constant.
The distinction between our proposed iterative update method and the existing approaches \cite{haoDeepKoopmanLearning2024} lies in the fact that we consider not only the addition of new data over time but also the \textbf{removal of outdated data}.
Therefore, the following three types of datasets are defined:
\begin{itemize}
    \item $\mathcal{S}_{\tau}^{\text{cur}} := \{x_k, u_k | k=k_{\tau}, k_{\tau}+1, ..., k_{\tau}+w-1 \}$ contains the \textit{current} snapshots involved in the calculation of the lifted system matrices at the $\tau$-th update, where $w$ is the window width. 
    \item $\mathcal{S}_{\tau}^{\text{new}} := \{x_k, u_k | k=k_{\tau}+w, k_{\tau}+w+1, ...\}$ contains the \textit{new} snapshots that have been collected after the $\tau$-th update but have not yet been utilized in the calculations.
    \item $\mathcal{S}_{\tau}^{\text{out}} := \{x_k, u_k | k=k_{\tau}-b, k_{\tau}-b+1, ..., k_{\tau}-1\}$ contains the \textit{outdated} snapshots that have been removed from the calculations at the $\tau$-th update, where $b$ is the batch size and $w\geq b$.
\end{itemize}
The evolution of the three datasets over time is illustrated in Fig.~\ref{fig:snapshots}.
The horizontal axis represents time, with the points along it from left to right corresponding to the snapshots obtained as the system evolves.
The blue points, yellow points, and red points outlined with a dashed box represent $\mathcal{S}_{\tau}^{\text{cur}},~\mathcal{S}_{\tau}^{\text{new}}, ~\mathcal{S}_{\tau}^{\text{out}}$, respectively.
The vertical axis from top to bottom shows the evolution of the three types of datasets and the updates of lifted system matrices.
Here, the number of data points contained in the dataset $\mathcal{S}_{\tau}^{\text{cur}}$ at each update index $\tau$ is set to be the same, defined as the window width $w$.
The window width $w$ and the update interval $b$ are tunable hyperparameters that need to be specified in advance.

To enable a more concise presentation in the following discussion, we first define the data matrices, as follows:
\begin{equation*}
    \begin{aligned}
        & X_{\tau}=[x_{k_{\tau}}, x_{k_{\tau}+1}, ..., x_{k_{\tau}+w-1}]\in \mathbb{R}^{n\times w},\\
        & Y_{\tau}=[x_{k_{\tau}+1}, x_{k_{\tau}+2}, ..., x_{k_{\tau}+w}]\in \mathbb{R}^{n\times w},\\
        & U_{\tau}=[u_{k_{\tau}}, u_{k_{\tau}+1}, ..., u_{k_{\tau}+w-1}]\in \mathbb{R}^{m\times w},\\
        & G_{\tau}=[g(x_{k_{\tau}}, \theta_{\tau}), ..., g(x_{k_{\tau}+w-1}, \theta_{\tau})]\in \mathbb{R}^{r\times w},\\
        & H_{\tau}=[g(x_{k_{\tau}+1}, \theta_{\tau}),  ..., g(x_{k_{\tau}+w}, \theta_{\tau})]\in \mathbb{R}^{r\times w}.\\
    \end{aligned}
\end{equation*}
Each column of these data matrices corresponds to one sampling instant, i.e., the data are stacked column-wise.
By minimizing the fitting error in (\ref{eq:koopman_state_hat}) based on the EDMD framework, one can obtain the optimal higher-dimensional system matrices:
\begin{align}
    \begin{bmatrix}
    A_{\tau} & B_{\tau}
    \end{bmatrix}
    &= H_{\tau}
    \begin{bmatrix}
    G_{\tau} \\
    U_{\tau}
    \end{bmatrix}^{\dagger},\label{eq:AB} \\
    C_{\tau}
    &= Y_{\tau} H_{\tau}^{\dagger}.\label{eq:C}
\end{align}
Since DNNs are used as the lifting functions, it is necessary to design a loss function to optimize the network parameters~$\theta$ via gradient descent.
By referring to \cite{haoDeepKoopmanLearning2024,luschDeepLearningUniversal2018,yeungLearningDeepNeural2019}, the loss function is defined as follows:
\begin{equation}\label{eq:loss}
    \begin{aligned}
        \theta_{\tau}^* = \arg\min\limits_{\theta_{\tau}} &\sum_{k=k_{\tau}}^{k_{\tau}+w-1} \{||x_{k+1} - C_{\tau}g(x_{k+1}, \theta_{\tau})||^2+\\
        &||g(x_{k+1}, \theta_{\tau}) -A_{\tau}g(x_{k}, \theta_{\tau})-B_{\tau}u_k||^2\}.
    \end{aligned}
\end{equation}
Further, the compact form of the loss function (\ref{eq:loss}) can be expressed as
\begin{equation}\label{eq:loss_compact}
    \begin{aligned}
        \theta_{\tau}^* = \arg\min\limits_{\theta_{\tau}} \{&||Y_{\tau} - C_{\tau}H_{\tau}||_F^2\\
        &+||H_{\tau} - A_{\tau}G_{\tau} - B_{\tau}U_{\tau}||_F^2\}.
    \end{aligned}
\end{equation}
We adopt the same strategy as in~\cite{haoDeepKoopmanLearning2024} to learn the initial system matrices and $\theta$. Specifically, the lifted system matrices and the neural network parameters are updated in an alternating manner, where one set of variables is held fixed while optimizing the other.
After a prescribed number of iterations, this process converges, yielding the initial higher-dimensional system matrices and lifting functions.
Subsequently, the higher-dimensional system matrices are updated online according to Proposition~\ref{update_rule}.
\begin{assumption}\label{assum:full_rank}
    \begin{rm}
        The matrices $\begin{bmatrix}
            G_{\tau}\\
           U_{\tau}
           \end{bmatrix}$ and $H_{\tau}$ are of full row rank.
    \end{rm}
\end{assumption}
\begin{remark}
    \begin{rm}
        The assumption of full row rank is commonly used in the Koopman-based methods~\cite{haseliGeneralizingDynamicMode2023,haseliLearningKoopmanEigenfunctions2022,haoDeepKoopmanLearning2024}.
        We remark that Assumption~\ref{assum:full_rank} is not a necessary condition for solving $A_{\tau},B_{\tau},C_{\tau}$ by~\eqref{eq:AB} and \eqref{eq:C}.
        Instead, this condition serves as a basis for the following iterative update algorithm.
        In Corollary~\ref{cor:condition}, we introduce a simplified condition to verify whether Assumption~1 holds at each update step.
        A necessary condition for the matrices $[G_{\tau}^{\top}\quad U_{\tau}^{\top}]^{\top} \in\mathbb{R}^{(r+m)\times w}$ in Assumption~\ref{assum:full_rank} to be of full row rank is that $w\geq r+m $, meaning that the number of snapshots in each $\mathcal{S}_{\tau}^{\text{cur}}$ must be sufficiently large.
    \end{rm}
\end{remark}
As shown in \cite{kordaConvergenceExtendedDynamic2018}, a necessary condition for the approximate Koopman operator obtained by EDMD to converge to the true Koopman operator is that $r\rightarrow \infty$.
This indirectly requires the window width $w$ to approach infinity, since $w > r+m$. This is unacceptable for time-varying systems, as obtaining the lifted system matrices would necessitate collecting a large number of snapshots, some of which may contain outdated data inconsistent with the current system dynamics.

To overcome these difficulties, we exploit the unique characteristics of deep neural networks when used as lifting functions. Here, the activation function after the final hidden layer is denoted as $\phi^h:=[\phi_1^h,\phi_2^h..., \phi_{n_h}^h]$, where $n_h$ is the number of neurons in the final hidden layer.
\begin{assumption}\label{assum:orthonormal}
    (Assumption~4 in \cite{haoDeepKoopmanLearning2024})
    \begin{rm}
        (1) Given a Hilbert space $\mathcal{F}$, the Koopman operator $\mathcal{K}$ is bounded and continuous on $\mathcal{F}$.
        (2) $\phi^h$ is selected from the orthonormal basis of $\mathcal{F}$, i.e., $[\phi_1^h,\phi_2^h..., \phi_{\infty}^h]$ is an orthonormal basis of $\mathcal{F}$.
    \end{rm}
\end{assumption}
\begin{lemma}\label{lem:convergence}
    (Lemma~3 in \cite{haoDeepKoopmanLearning2024}):
    \begin{rm}
        If Assumption~\ref{assum:orthonormal} holds, then the approximate Koopman operator converges to the true Koopman operator $\mathcal{K}$ as $n_h\rightarrow \infty$.
    \end{rm}
\end{lemma}
Based on Lemma~\ref{lem:convergence}, the necessary condition $r\rightarrow\infty$ is replaced by $n_h\rightarrow\infty$, when employing DNNs as observables. 
This allows the window width \(w\) to be reduced, resulting in a decrease in the number of snapshots included in $\mathcal{S}_{\tau}^{\text{cur}}$. 
By using DNNs as observables, it is unnecessary to include a large amount of outdated data, and the lifted higher-dimensional linear system is aligned with the present nonlinear system.
    
\subsection{Iterative update of lifted system matrices}
The neural network structure employed in this paper is a multilayer perceptron (MLP) and the detailed layer-wise architecture will be provided in Section~\ref{sec:simulation} for the numerical simulation examples.
When using DNNs as lifting functions, the optimal network weight parameters \( \theta \) are obtained by performing gradient descent on the loss function (\ref{eq:loss_compact}). 
Since the gradient descent process is time-consuming, a certain amount of input-output data must be collected offline to enable offline training of the network parameters.
We train the weights of the DNN, which serves as the lifting function, using a small amount of input-output data collected during the early operation of the time-varying system.

Throughout the system online operation, the dataset $\mathcal{S}_{\tau}^{\text{new}}$ is continuously augmented with new snapshots as time progresses.
The lifted system matrices will be updated once the number of newly collected snapshots reaches a specified threshold (batch size $b$).
If each update requires reconstructing $\mathcal{S}_{\tau+1}^{\text{c}}$ and repeating the calculation of the lifted system matrices as shown in (\ref{eq:AB}) and (\ref{eq:C}), it would lead to a significant waste of computational time.
Therefore, an iterative method for updating lifted system matrices is proposed, as follows. For convenience, denote $g_{k}:=g(x_{k}, \theta)$.
\begin{proposition}\label{update_rule}
    (The update rule of the lifted system matrix):
    \begin{rm}
        If the matrices $[G_{\tau}^{\top}\quad U_{\tau}^{\top}]^{\top},H_{\tau},[G_{\tau+1}^{\top}\quad U_{\tau+1}^{\top}]^{\top}$, $H_{\tau+1}$ are of full row rank (Assumption~\ref{assum:full_rank} holds at the update indices $\tau \text{ and }\tau+1$), and the matrices $E+Z_{\tau}^{\top}P_{\tau}Z_{\tau}, E+W_{\tau}^{\top}\overline{P}_{\tau}W_{\tau}$ defined below are invertible, then the lifted system matrices can be updated iteratively based on $\mathcal{S}_{\tau}^{\text{cur}} \text{ and } \mathcal{S}_{\tau}^{\text{new}}$, as follows:
        \begin{equation}\label{eq:AB_iteration}
            \begin{aligned}
                \begin{bmatrix}
                    A_{\tau+1} & B_{\tau+1}
                \end{bmatrix}=&
                \begin{bmatrix}
                    A_{\tau} & B_{\tau}
                \end{bmatrix}\\
                &+(W_{\tau}-\begin{bmatrix}
                    A_{\tau} & B_{\tau}
                    \end{bmatrix}Z_{\tau})\Gamma_{\tau}Z_{\tau}^{\top}P_{\tau},
            \end{aligned}
        \end{equation}
        \vspace{-0.7cm}
        \begin{align}
            P_{\tau+1}&=P_{\tau}-P_{\tau}Z_{\tau}\Gamma_{\tau}Z_{\tau}^{\top}P_{\tau}\label{eq:P},\\
            C_{\tau+1}&=C_{\tau}+(V_{\tau}-C_{\tau}W_{\tau})\overline{\Gamma}_{\tau}W_{\tau}^{\top}\overline{P}_{\tau}\label{eq:C_iteration},\\
            \overline{P}_{\tau+1}&=\overline{P}_{\tau}-\overline{P}_{\tau}W_{\tau}\overline{\Gamma}_{\tau}W_{\tau}^{\top}\overline{P}_{\tau},
        \end{align}
        where 
        \begin{equation*}
            \begin{aligned}
                &Z_{\tau}=\begin{bmatrix}
                    g_{k_{\tau}}&...&g_{k_{\tau}+b-1}&g_{k_{\tau}+w}&...&g_{k_{\tau}+w+b-1}\\
                    u_{k_{\tau}} & ... & u_{k_{\tau}+b-1} & u_{k_{\tau}+w} & ... & u_{k_{\tau}+w+b-1}\\
                \end{bmatrix},\\
                &W_{\tau}=\begin{bmatrix}
                    g_{k_{\tau}+1} & ... & g_{k_{\tau}+b} & g_{k_{\tau}+w+1} & ... & g_{k_{\tau}+w+b} \\
                \end{bmatrix},\\
                &V_{\tau}=\begin{bmatrix}
                    x_{k_{\tau}+1} & ... & x_{k_{\tau}+b} & x_{k_{\tau}+w+1} & ... & x_{k_{\tau}+w+b}\\
                \end{bmatrix},
            \end{aligned}
        \end{equation*}
        $\Gamma_{\tau}=(E+Z_{\tau}^{\top}P_{\tau}Z_{\tau})^{-1},\overline{\Gamma}_{\tau}=(E+W_{\tau}^{\top}\overline{P}_{\tau}W_{\tau})^{-1},E=\begin{bmatrix}
                    -I_{b}& \\
                    &I_b
                \end{bmatrix}$ need to be reconstructed and calculated at every instant.
        The initial matrix values $P_{0}=\left(\begin{bmatrix}
            G_{0} \\
            U_{0}
            \end{bmatrix}\begin{bmatrix}
                G_{0} \\
                U_{0}
                \end{bmatrix}^{\top}\right)^{-1}$, $\overline{P}_{0}=(H_{0}H_{0}^{\top})^{-1}$ are set at the initial stage and are used to update iteratively.
    \end{rm}
\end{proposition}

The proof of Proposition \ref{update_rule} is given in Appendix \ref{apdx:A}.
There is an intuitive interpretation for the update method (\ref{eq:AB_iteration}). 
The quantity $(W_{\tau}-\begin{bmatrix}
    A_{\tau} & B_{\tau}
    \end{bmatrix}Z_{\tau})$ can be regarded as the prediction error associated with the current lifted system, specifically $A_{\tau}$ and $B_{\tau}$. 
Then, the lifted system matrices are updated by incorporating a term that is proportional to this error.
Since each update of the lifted system matrices discards outdated data, the resulting data matrices \( [G_{\tau+1}^{\top}\quad U_{\tau+1}^{\top}]^{\top} \) and \( H_{\tau+1} \) may no longer satisfy Assumption~\ref{assum:full_rank}.
Therefore, the condition is required to verify at each update instant.

\begin{remark}
    (Time complexity of the update rule):
    \begin{rm}
        {The total time complexity for updating the matrices based on the datasets \(S_{\tau}^{\text{new}}\) and \(S_{\tau}^{\text{out}}\) using Proposition~\ref{update_rule} is \(\mathcal{O}(b r^{2})\).
        If each update recomputes the pseudoinverse using formula~\eqref{eq:AB} based on the dataset \(S_{\tau}^{\text{cur}}\), the time complexity becomes \(\mathcal{O}(w r^{2})\), where \(w > b\).
        This is consistent with the computational complexity results for windowed DMD reported in~\cite{zhangOnlineDynamicMode2019}.
        Comparing the two approaches, it is evident that the method in Proposition~\ref{update_rule} substantially reduces the computational load, making it more suitable for online updates.}
    \end{rm}
\end{remark}

Due to the removal of outdated data, the preconditions given in Proposition \ref{update_rule} must be verified to ensure feasibility, which takes considerable computational time.
In order to simplify the verification of the required conditions, we make use of the following equivalence.

\begin{lemma}\label{update_condition}
    \begin{rm}
        If $[G_{\tau}^{\top}\quad U_{\tau}^{\top}]^{\top}$ and $H_{\tau}$ are of full row rank, then $[G_{\tau+1}^{\top}\quad U_{\tau+1}^{\top}]^{\top}$ and $H_{\tau+1}$ are of full row rank if and only if $E+Z_{\tau}^{\top}P_{\tau}Z_{\tau}$ and $E+W_{\tau}^{\top}\overline{P}_{\tau}W_{\tau}$ are invertible, where $E, Z_{\tau}, W_{\tau}, P_{\tau} \text{ and } \overline{P}_{\tau}$ are defined in Proposition~\ref{update_rule}.
    \end{rm}
\end{lemma}
The proof of Lemma~\ref{update_condition} is given in Appendix \ref{apdx:B}.
Lemma~\ref{update_condition} shows the equivalence between that $[G_{\tau+1}^{\top}\quad U_{\tau+1}^{\top}]^{\top}$, $H_{\tau+1}$ are of full row rank (Assumption~\ref{assum:full_rank} holds at the $(\tau+1)$-th update) and that $E+Z_{\tau}^{\top}P_{\tau}Z_{\tau}$ and $E+W_{\tau}^{\top}\overline{P}_{\tau}W_{\tau}$ are invertible.
Therefore, increasing $w$ is beneficial for ensuring that the above data matrices are full rank.
Based on Lemma~\ref{update_condition}, the feasibility conditions required in Proposition~\ref{update_rule} can be simplified.
\begin{corollary}\label{cor:condition}
    (Efficient Online Feasibility Check):
    \begin{rm}
The conditions required in Proposition~1 can be simplified to verifying only the invertibility of the low-dimensional matrices \(E + Z_{\tau}^{\top} P_{\tau} Z_{\tau}\) and \(E + W_{\tau}^{\top} \overline{P}_{\tau} W_{\tau}\).
    \end{rm}
\end{corollary}
Corollary~\ref{cor:condition} indicates that the time complexity for verifying feasibility conditions required in Proposition~\ref{update_rule} is reduced from $\mathcal{O}(r^3+b^3)$ to $\mathcal{O}(b^3)$, making the method more suitable for online processing.

\subsection{Estimation error analysis}
To quantify the estimation error of the aforementioned Koopman operator, we make the following assumptions.
\begin{assumption}\label{assum:time-varying}
    \begin{rm}
        The time interval $\Delta t =t_{k+1}-t_k$ is sufficiently small such that there exist $\mu_x \geq 0$ and $\mu_u \geq 0$, satisfying $\|x_{k+1}-x_k\| \leq \mu_x < \infty$ and $\|u_{k+1}-u_k\| \leq \mu_u < \infty$.    
    \end{rm}
\end{assumption}
This assumption is always satisfied for systems with bounded state variations.
It delineates the class of nonlinear time-varying systems for which our algorithm is applicable.
We introduce a common assumption posed in neural network theory.
\begin{assumption}\label{assum:Lipschitz}
    (Assumption~3 in~\cite{haoDeepKoopmanLearning2024})
    \begin{rm}
        The observable function DNN $g(\cdot, \theta)$ is Lipschitz continuous on the system state space with Lipschitz constant $\mu_g$.
    \end{rm}
\end{assumption}
We define the estimation error of the Koopman operator as $e_k=x_k-\hat{x}_k$, where $\hat{x}_k$ is the predicted state obtained by (\ref{eq:koopman_state_hat}).
\begin{lemma}
\label{lem:estimation_error}
    (Estimation error of the Koopman operator):
    \begin{rm}
        If Assumptions~\ref{assum:full_rank}, \ref{assum:orthonormal}, \ref{assum:time-varying} and \ref{assum:Lipschitz} hold, then the estimation error of the Koopman operator satisfies
        \begin{equation}\label{eq:estimation_error}
            \lim_{n_h\rightarrow\infty}\sup \|e_k\| \leq (\|C_{\tau}A_{\tau}\|\mu_g+1)\mu_x+\|C_{\tau}B_{\tau}\|\mu_u+E_{\text{recon}},
        \end{equation}
        {where $\mu_x, \mu_u,$ and $\mu_g$ are the time-varying rates and the Lipschitz constants of the neural network defined in Assumptions~\ref{assum:time-varying} and~\ref{assum:Lipschitz}, respectively, and $E_{\text{recon}}$ is the reconstruction error, i.e., $E_{\text{recon}} :=\max_{x_k\in\mathcal{S}_{\tau}^\text{cur}}\|x_k-C_{\tau}g(x_k,\theta_{\tau})\|$.}
    \end{rm}
\end{lemma}
The proof of Lemma~\ref{lem:estimation_error} is similar to Theorem~1 in \cite{haoDeepKoopmanLearning2024} and given in Appendix~\ref{apdx:error}.
{Equation~\eqref{eq:estimation_error} provides a theoretical upper bound for the estimation error in the asymptotic limit of infinite network width ($n_h \to \infty$). 
Evidently, the estimation error depends on the system's variation rates ($\mu_x, \mu_u$), the network's Lipschitz property ($\mu_g$), and the reconstruction capability ($E_{\text{recon}}$).}
In practical applications, it is impossible to construct a DNN with an infinite number of hidden neurons. Therefore, an additional loss term is incorporated during neural network training to ensure $\|A_{\tau}\|_2 < 1$.
Under this condition, an upper bound for the estimation error as $k \to \infty$ can be derived, which depends on $\mu_x$, $\mu_u$, and the neural network fitting error presented in (\ref{eq:loss}).\footnote{Corollary 1 in \cite{haoDeepKoopmanLearning2024}: if $\|A_{\tau}\|_2<1$ and Assumptions~\ref{assum:full_rank},\ref{assum:time-varying},\ref{assum:Lipschitz} hold, then the upper bound of $\|e_k\|$ is determined by the minimization performance of (\ref{eq:loss}) and $\mu_x,\mu_u$ as $k\rightarrow\infty$.}

\subsection{Update mechanisms}
Compared to the results in \cite{haoDeepKoopmanLearning2024}, our proposed method only uses the most recent data to fit the higher-dimensional linear system.
However, this leads to the following question: \emph{For time-varying systems, does incorporating new data and discarding outdated data to fit the lifted system matrices necessarily reduce the prediction error}?

Note that while the proposed method targets nonlinear systems, here we employ a simple linear time-varying system solely as a motivating example.
Consider the following time-varying linear system:
\begin{equation}
    \begin{bmatrix}
        \dot{x}(t) \\
        \dot{y}(t)
    \end{bmatrix}=\begin{bmatrix}
        0 & 1\\
        -1 & -0.2+0.5\sin(t)
    \end{bmatrix}\begin{bmatrix}
        x(t)\\
        y(t)
    \end{bmatrix},
\end{equation}
where $x$ and $y$ are the state variables, and $t$ is the time variable.
This continuous-time system is converted to a discrete-time system using the Runge-Kutta method with a given time interval $\delta_t=0.1s$, and a 10s trajectory, i.e., 100 snapshots, is simulated starting from the initial state $[x(0),y(0)]^{\top}=[1,1]^{\top}$.

\begin{figure}
    \begin{center}
    \includegraphics[height=7cm]{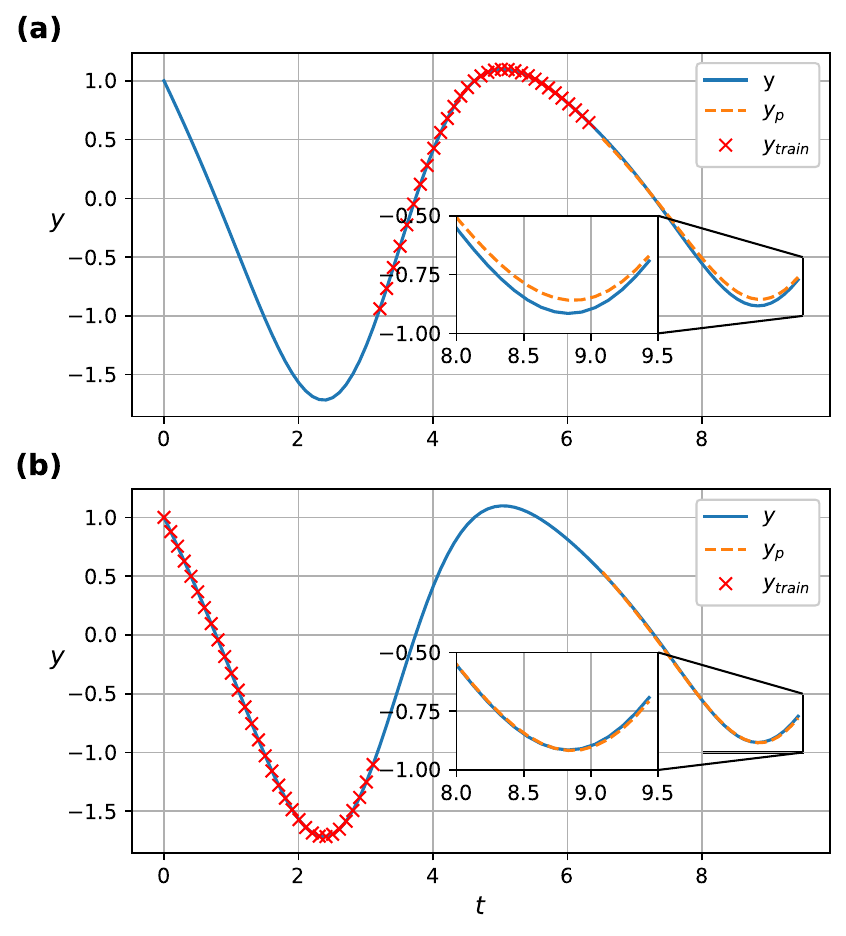}  
    \caption{Comparison of prediction performance using new or outdated data.
    In the figure, the blue solid line represents the true trajectory of the system state \(y\), the red dots indicate the snapshots used for fitting the system, and the yellow dashed line represents the predicted system state. 
    A part of the trajectory is enlarged for clearer visualization.}  
    \label{fig:newVSold}     
    \end{center}     
\end{figure}
The simulation trajectory of the system state \(y\) is shown in Fig.~\ref{fig:newVSold}. 
Figs.~\ref{fig:newVSold}(a) and (b) illustrate the prediction results using new data and outdated data, respectively.
Comparing these two figures, it can be seen that using new data to fit the system results in a decline in short-term prediction performance (as shown in the magnified sections of the figures).
Here, a metric for evaluating prediction error, namely the Root Mean Square Error (RMSE), is used:
\begin{equation}
    RMSE:=\sqrt{\frac{1}{N}\sum_{i=1}^{N}(y_{\text{true}}(i)-y_{\text{pred}}(i))^2}.
\end{equation}
The RMSE for the prediction results over 30 sampling time steps using outdated data and new data for fitting the system are 0.0050 and 0.0274, respectively.
Therefore, for some specific systems, using new data to fit the system does not necessarily yield better short-term prediction results compared to the use of outdated data. 
This may be due to the periodic nature of the system.

Based on the above observations, a result-oriented update mechanism is designed to determine whether an update is necessary, aiming to enhance predictive performance for general systems.
New data will be incorporated into fitting the lifted system matrices when the following condition is satisfied:
\begin{equation}\label{eq:event_trigger}
    \begin{aligned}
        &\left\|C_{\tau+1}\begin{bmatrix}
            A_{\tau+1} & B_{\tau+1}
        \end{bmatrix}\begin{bmatrix}
            g(X_{\tau+1}^{\text{new}}, \theta_{\tau+1})\\
            U_{\tau+1}^{\text{new}}
        \end{bmatrix}-Y_{\tau+1}^{\text{new}}\right\|_F^2\\
        &\leq\left\|C_{\tau}\begin{bmatrix}
            A_{\tau} & B_{\tau}
        \end{bmatrix}\begin{bmatrix}
            g(X_{\tau+1}^{\text{new}}, \theta_{\tau})\\
            U_{\tau+1}^{\text{new}}
        \end{bmatrix}-Y_{\tau+1}^{\text{new}}\right\|_F^2,
    \end{aligned}
\end{equation}
where $A_{\tau+1}, B_{\tau+1}\text{ and }C_{\tau+1}$ are the updated lifted system matrices computed by Proposition \ref{update_rule}, and $X_{\tau+1}^{\text{new}},Y_{\tau+1}^{\text{new}}$ and $U_{\tau+1}^{\text{new}}$ are the new snapshots in $\mathcal{S}_{\tau}^{\text{new}}$.
An intuitive explanation of this update mechanism is as follows: after constructing a new surrogate model using the most recent online data points in $\mathcal{S}_{\tau}^\text{new}$, if its fitting error on the new dataset exceeds that of the previous surrogate model, the update is rejected and the previous model is reused.

It should be noted that verifying this condition requires significant computational time.
Therefore, an additional condition is introduced to reduce unnecessary updates. 
Specifically, the condition in (\ref{eq:event_trigger}) is no longer verified, and the lifted system matrices will not be updated when the following condition is satisfied:
\begin{equation}\label{eq:event_trigger2}
    \left\|C_{\tau}\begin{bmatrix}
        A_{\tau} & B_{\tau}
    \end{bmatrix}\begin{bmatrix}
        g(X_{\tau+1}^{\text{new}}, \theta_{\tau})\\
        U_{\tau+1}^{\text{new}}
    \end{bmatrix}-Y_{\tau+1}^{\text{new}}\right\|_F^2\leq \epsilon,
\end{equation}
where $\epsilon\geq 0$ is a predefined threshold.
An intuitive interpretation of (\ref{eq:event_trigger2}) is that if the fitting error of the previous surrogate model on the new data points in $\mathcal{S}_{\tau}^\text{new}$ is smaller than a predefined threshold~$\epsilon$, the model is considered to be reliable, hence no new surrogate model is computed, thereby saving computational time.
The choice of \( \epsilon \) can be guided by the approximation error obtained during the initial training phase.

This online learning algorithm for unknown time-varying nonlinear systems is presented as Algorithm~\ref{alg:online_learning}.
{In the initial training stage, we employ an alternating minimization strategy to obtain the initial lifted system matrices and the DNN parameters \( \theta \).}
Whenever the number of new data points in the dataset \(\mathcal{S}_{\tau}^\text{new}\) meets the required amount, condition~(\ref{eq:event_trigger2}) is first evaluated to determine whether the previous model remains reliable.
The proposed algorithm operates with a fixed memory budget. The sliding window mechanism ensures that the number of retained snapshots is bounded by $w + b$, where $w$ is the window size and $b$ is the batch size.
Outdated snapshots are discarded to enable the surrogate model to track time-varying dynamics.

\begin{algorithm}
    \caption{Online learning algorithm for unknown time-varying nonlinear systems}\label{alg:online_learning}
    \begin{algorithmic}[1]
        \Require $w,~b$, the initial $w$ snapshots, the predefined threshold $\epsilon$, the structure of the neural network used as the lifting function and its training parameters.
        \Ensure $A_{\tau}, B_{\tau}, C_{\tau}, \theta_{\tau}$, and the prediction of the state.
        \State Use gradient descent on the initial snapshots to train the DNN and obtain the network weights $\theta_0$.
        Initialize $A_{0}, B_{0}, C_{0}$ by (\ref{eq:AB}), (\ref{eq:C}) and calculate $P$ and $\overline{P}$, $\tau \gets 0$.
        \While{the system is running}
        \If{$\mathcal{S}_{\tau}^{\text{new}}$ contains $b$ new snapshots and the condition in Corollary~\ref{cor:condition} is satisfied}
        \State Calculate the fitting error of the current model $(A_\tau, B_\tau)$ on $\mathcal{S}_{\tau}^{\text{new}}$.
        \If { error $\le \epsilon$ (Condition~\eqref{eq:event_trigger2} satisfied)}
        \State Clear $\mathcal{S}_{\tau}^{\text{new}}$.
        \ElsIf {the condition in (\ref{eq:event_trigger}) is satisfied}
        \State Update $A_{\tau+1}, B_{\tau+1}, C_{\tau+1}$ by Prop. \ref{update_rule}.
        \State Update $\theta_{\tau+1}$ by solving (\ref{eq:loss_compact}).
        \State Remove $\mathcal{S}_{\tau}^{\text{out}}$ from $\mathcal{S}_{\tau}^{\text{cur}}$; Add $\mathcal{S}_{\tau}^{\text{new}}$ to $\mathcal{S}_{\tau}^{\text{cur}}$; Clear $\mathcal{S}_{\tau}^{\text{new}}$.
        \EndIf
        \Else 
        \State Given $x_k$ and $u_k$ of the original system, $\hat{x}_{k+1}$ is obtained based on (\ref{eq:koopman_state_hat}).
        \State $\mathcal{S}_{\tau}^{\text{new}} \leftarrow \mathcal{S}_{\tau}^{\text{new}} \cup \{(x_k, u_k)\},k\leftarrow k+1$.
        \EndIf
        \EndWhile
    \end{algorithmic}
\end{algorithm}

\section{ Stability-guaranteed controller design}\label{sec:controller}
\vspace{-0.3cm}
The previous section introduced an online iterative update algorithm for nonlinear time-varying systems using the deep Koopman operator.
{This section focuses on controller design based on the surrogate model~\eqref{eq:koopman_state_hat} obtained by Algorithm~\ref{alg:online_learning} to guarantee the input-to-state stability of the closed-loop system.}
\subsection{Controller design}
\begin{figure*}
    \centering
    \includegraphics[width=0.9\textwidth]{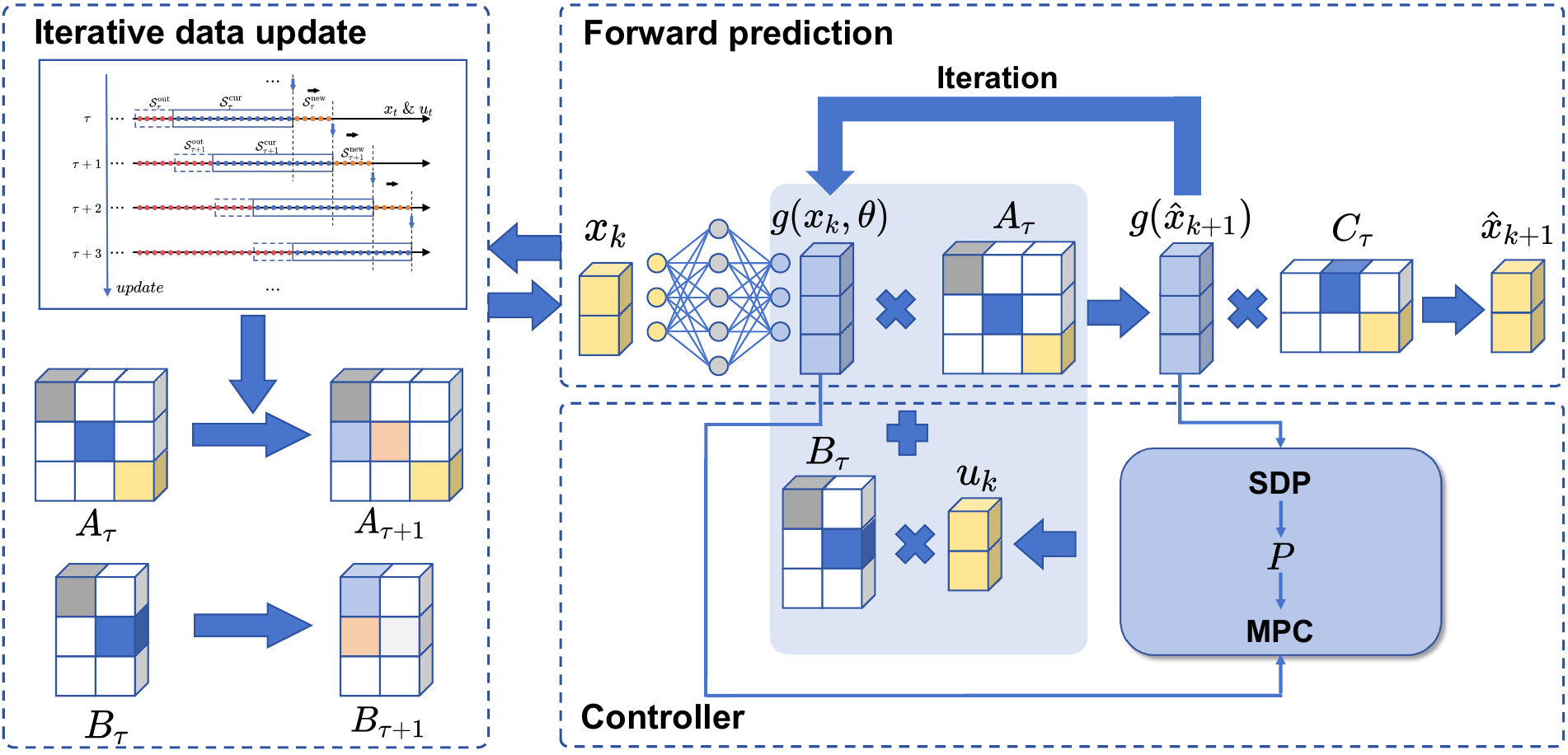}
    \caption{{Control framework for unknown time-varying nonlinear systems.
    The left part of the figure, \emph{Data update iteratively}, illustrates that the lifted system matrices are updated online based on the measured trajectories (where the color change reflects the variation in matrix entries).
    The right part, \emph{Forward prediction}, shows the iterative forward prediction of the unknown nonlinear system state using the lifted linear model.
    During trajectory tracking, the \emph{Controller} module computes the control input \(u_k\).}
    }\label{fig:control_framework}
\end{figure*}

\vspace{-0.2cm}
Here, \( g(x_k,\theta_\tau) \) denotes the state of the exact high-dimensional linear system as follows:
\begin{equation}\label{eq:true_state}
    g(x_{k+1}, \theta_\tau) = A_{\tau}g(x_{k},\theta_\tau) + B_{\tau}u_{k} + \epsilon_{k},
\end{equation}
{where $x_{k+1}=f(x_{k},u_{k},k)$ based on~\eqref{eq:system}, and $\epsilon_{k}$ denotes the deterministic modeling residual resulting from the approximation of the nonlinear dynamics, which is determined by the loss function in (\ref{eq:loss}).}
In the above, $\hat{g}_k\in\mathbb{R}^r$ is the nominal system state.
The nominal system is described by the linear time-varying system:
\begin{equation}\label{eq:nominal_system}
    \begin{aligned}
        \hat{g}_{k+1} &= A_{\tau}\hat{g}_k + B_{\tau}u_k,\\
        \hat{x}_k &= C_{\tau}\hat{g}_k, \hat{g}_{0} = g(x_0, \theta),\\
    \end{aligned}    
\end{equation}
where $\mathbb{U}\subseteq\mathbb{R}^m$ is the control input space and $k,\tau\in\mathbb{Z}_{\geq 0}$ are the time index and the update index, respectively.
Inspired by\cite{chen2026data,chen2024learning}, we design an MPC controller based on the nominal system. The control input is obtained by solving the following optimization problem:
\begin{equation}\label{eq:mpc}
    \begin{aligned}
        \min_{u_{k:k+H-1,k}}&\sum_{j=0}^{H-1}(\|\hat{g}_{k+j,k}\|_Q^2+\|u_{k+j,k}\|_R^2)+\|\hat{g}_{k+H,k}\|_{P}^2\\
        \text{s.t. } &\hat{g}_{k+j+1,k} = A_{\tau}\hat{g}_{k+j,k}+B_{\tau}u_{k+j,k},\\
        &\hat{g}_{k,k} = g(x_k,\theta_\tau),u_k\in\mathbb{U},\\
    \end{aligned}
\end{equation}
where $H$ is a constant prediction horizon, $Q=(Q^{\frac{1}{2}})^{\top}(Q^{\frac{1}{2}})\in\mathbb{R}^{n\times n}$, $R\in\mathbb{R}^{m\times m}$, and $P\in\mathbb{R}^{r\times r}$ are the positive definite weighting matrices for the state, control input, and terminal state, respectively.
Moreover, $\hat{g}_{k+j,k}$ denotes the predicted state at time $k+j$ based on the current state $\hat{g}_k$ and the control input sequence $u_{k:k+j-1,k}$.
The minimum of the above cost function is defined as the optimal value function, denoted by $V^*(k,\hat{g}_k)$.
The first element of the optimal control sequence corresponds to the optimal control input at time instant $k$, denoted by $u_k^*$, which is applied to the nonlinear system.
The control input is computed in the lifted space, which is not decoded back to the physical state.
To ensure the recursive feasibility of the above MPC problem and the stability of the resulting closed-loop system, the positive definite matrix $P$ must be carefully designed so that it satisfies the following properties:
\begin{equation}\label{eq:P_matrix}
    \begin{aligned}
        &\hat{g}_{k,k}^{\top}P\hat{g}_{k,k}\leq \gamma,\\
        &P -(A_{\tau}+B_{\tau}K)^{\top}P(A_{\tau}+B_{\tau}K) - Q - K^{\top}RK \succeq 0,\\
    \end{aligned}
\end{equation}
where $K$ refers to the feedback gain matrix, $\gamma\geq 0$ is the size of the ellipsoid invariant set and $Q$, $R$ are the weighted matrices.
However, in the above formulation, the matrices $P$ and $K$ are coupled. Inspired by~\cite{kothare1996robust,chen2026data}, we apply a variable substitution to decouple them by setting $\overline{P}=\gamma P^{-1}$ and $Y= K\overline{P}$.
Then, the problem of solving $\gamma$ and the matrix $P$ in~\eqref{eq:P_matrix} can be reformulated as the following semidefinite programming (SDP) problem:
\vspace{-0.8cm}
{\small \begin{equation}\label{eq:SDP}
    \begin{aligned}
        &\min_{\gamma,\overline{P},Y}\quad\gamma,\\
        \text{s.t. }&\begin{bmatrix}
            \overline{P}&(A_\tau\overline{P}+B_\tau Y)^\top&(Q^{\frac{1}{2}}\overline{P})^\top&(R^{\frac{1}{2}}Y)^\top\\
            A_\tau\overline{P}+B_\tau Y&\overline{P}&0&0\\
            Q^{\frac{1}{2}}\overline{P}&0&\gamma I&0\\
            R^{\frac{1}{2}}Y&0&0&\gamma I
        \end{bmatrix}\succeq 0,\\
        &\begin{bmatrix}
            1&\hat{g}_{k,k}^\top\\
            \hat{g}_{k,k}&\overline{P}
        \end{bmatrix}\succeq 0,\begin{bmatrix}
            \overline{U}&Y\\
            Y^\top  &\overline{P}
        \end{bmatrix}\succeq 0,\overline{U}=\text{Diag}(u_{\max}),
    \end{aligned}
\end{equation}}
where $\overline{U} = \mathrm{Diag}(u_{\max})$ is a diagonal matrix whose diagonal entries are given by $u_{\max}$, representing the maximum control input.
The first two semidefinite matrix constraints are derived from~\eqref{eq:P_matrix}, while the last semidefinite constraint enforces that each component of the control input $u$ does not exceed the prescribed bound $u_{\max}$.
Notably, all constraints in the SDP are linear matrix inequalities, making it a convex optimization problem that can be solved efficiently online.
At each time step, the SDP~\eqref{eq:SDP} is solved to obtain the terminal cost weighting matrix $P$, after which the MPC problem~\eqref{eq:mpc} is solved to compute the control input $u_{k,k}^*$, which is then applied to the original nonlinear system.
The detailed algorithm is provided in Algorithm~\ref{alg:online_control}.
The control framework is illustrated in Fig~\ref{fig:control_framework}.
\begin{algorithm}[t]
    \caption{Online iterative updating and control algorithm for unknown time-varying nonlinear systems}\label{alg:online_control}
    \begin{algorithmic}[1]
        \Require $w,~b$, the initial $w$ snapshots, the predefined threshold $\epsilon$, the structure of the neural network used as the lifting function and its training parameters, MPC parameters and the desired trajectory.
        \Ensure The control input at each time index.
        \State Use gradient descent on the initial snapshots to train the DNN and obtain the network weights $\theta$.
        Initialize $A_{0}, B_{0}, C_{0}$ by (\ref{eq:AB}), (\ref{eq:C}) and calculate $P$ and $\overline{P}$, $\tau \gets 0$.
        \While{the system is running}
        \If{$\mathcal{S}_{\tau}^{\text{new}}$ contains $b$ new snapshots and the condition in Corollary~\ref{cor:condition} is satisfied}
        \State {Calculate the fitting error of the current model $(A_\tau, B_\tau)$ on $\mathcal{S}_{\tau}^{\text{new}}$.}
        \If { error $\le \epsilon$ (Condition~\eqref{eq:event_trigger2} satisfied)}
        \State Clear $\mathcal{S}_{\tau}^{\text{new}}$.
        \ElsIf {the condition in (\ref{eq:event_trigger}) is satisfied}
        \State Update $A_{\tau+1}, B_{\tau+1}, C_{\tau+1}$ by Prop. \ref{update_rule}.
        \State Update $\theta_{\tau+1}$ by solving (\ref{eq:loss_compact}).
        \State {Remove $\mathcal{S}_{\tau}^{\text{out}}$ from $\mathcal{S}_{\tau}^{\text{cur}}$; Add $\mathcal{S}_{\tau}^{\text{new}}$ to $\mathcal{S}_{\tau}^{\text{cur}}$; Clear $\mathcal{S}_{\tau}^{\text{new}}$.}
        \EndIf        
        \Else 
        \State Solve the SDP~\eqref{eq:SDP} to compute $P$.
        \State Solve the MPC problem~(\ref{eq:mpc}) based on $g(x_k,\theta)$, and denote the first element of the optimal control sequence as \( u_k^* \).
        \State Apply \( u_k^* \) to the actual nonlinear system.
        \State Add new snapshot to $\mathcal{S}_{\tau}^{\text{new}}$, $k\leftarrow k+1$.
        \EndIf
        \EndWhile
    \end{algorithmic}
\end{algorithm}

\subsection{Stability analysis}
\begin{lemma}\label{thm:RoA}
    \begin{rm}
        The set $\Xi:=\left\{\hat{g}\in\mathbb{R}^r|\hat{g}^\top P\hat{g}\leq \gamma\right\}$ is a domain of attraction for the nominal system $\hat{g}_{k+1}=A_{\tau}\hat{g}_k+B_{\tau}u_k$ under the MPC controller~\eqref{eq:mpc}.
    \end{rm}
\end{lemma}
The proof of Lemma~\ref{thm:RoA} is given in Appendix \ref{apdx:E}.
To ensure the recursive feasibility of the MPC scheme and the existence of a feasible solution to the SDP problem, we introduce the following assumption.
\begin{assumption}\label{assum:gamma}
    (Assumption~8 in~\cite{chen2024learning}; Assumption~6 in~\cite{chen2026data})
    \begin{rm}
        The upper bound $\gamma$ satisfies $\gamma-{g}_k^{\top}P{g}_k\geq\mu_k\left\|\epsilon_k\right\|,\forall k \in\mathbb{Z}_{\geq 0}$, where $\mu_k$ is a constant satisfying $\mu_k\geq\|[g(f(x_k,Kg_k))+(A_{\tau}+B_{\tau}K)g_k]^{\top}P\|$.
    \end{rm}
\end{assumption}
Assumption~\ref{assum:gamma} implies that the current high-dimensional state stays adequately away from the boundary, leaving enough margin to fully absorb the impact caused by the modeling error $\epsilon_k$.
This requires the modeling accuracy to be sufficiently high; therefore, the high-dimensional system matrices must be iteratively updated according to Proposition~\ref{update_rule}.
\begin{theorem}\label{thm:recursive_feasibility}
    \begin{rm}
        Suppose that the SDP~\eqref{eq:SDP} for computing the terminal cost and the MPC problem~\eqref{eq:mpc} are feasible at time instant $k$.
        If Assumption~\ref{assum:gamma} holds, then the recursive feasibility of the MPC problem~\eqref{eq:mpc} and the existence of a feasible solution to the SDP problem~\eqref{eq:SDP} at time instant $k+1$ are guaranteed.
    \end{rm}
\end{theorem}
The proof of Theorem~\ref{thm:recursive_feasibility} is given in Appendix \ref{apdx:F}.
After establishing the feasibility of the SDP problem at each time instant and the recursive feasibility of the MPC problem, we can characterize the stability properties of the resulting closed-loop system. First, the following lemma is introduced.
\begin{lemma}
    (Theorem~1 in~\cite{li2018input})
    \begin{rm}
        Consider the system~\eqref{eq:system} with the MPC controller~\eqref{eq:mpc}. The closed-loop system is input-to-state stable if there exist three class $\mathscr{K}_\infty$ functions $\beta_1(\cdot)$, $\beta_2(\cdot)$, $\beta_3(\cdot)$, a class $\mathscr{K}$ function $\sigma(\cdot)$ and a function $V(k,x):\mathbb{Z}_{\geq0}\times\mathbb{R}^n\rightarrow\mathbb{R}_{\geq0}$ such that
        \begin{equation}
            \begin{aligned}
                \beta_1(\|g_k\|)\leq& V(k,g_k)\leq \beta_2(\|g_k\|),\\
                V(k+1,g_{k+1})-&V(k,g_k)\leq -\beta_3(\|g_k\|)+\sigma(\|\epsilon_k\|).
            \end{aligned}
        \end{equation}
        The function $V(k,g_k)$ is called an ISS Lyapunov function.
    \end{rm}
\end{lemma}
With this lemma in place, we can prove that the lifted high-dimensional system satisfies input-to-state stability.
\begin{theorem}\label{thm:ISS}
    \begin{rm}
        Suppose Assumption~\ref{assum:gamma} holds. The lifted high-dimensional system under the proposed MPC controller~\eqref{eq:mpc} is input-to-state stable with respect to the approximation error $\epsilon_k$.
    \end{rm}
\end{theorem}
The proof of Theorem~\ref{thm:ISS} is given in Appendix \ref{apdx:G}.
As is common in many Koopman-based methods~\cite{shiDeepKoopmanOperator2022,jiaEVOLVEROnlineLearning2024,kordaLinearPredictorsNonlinear2018}, the lifting function can be designed as a concatenation of the original nonlinear system state and the output of a DNN. In this case, $C_\tau = [I_n\; 0_{n\times r}]$.
Since the relationship between the original system state and the Koopman lifted state is given by $x_k = C_\tau g(x_k,\theta_\tau)$, the input-to-state stability of the closed-loop system implies that the original system state is also input-to-state stable.
\begin{corollary}
    \begin{rm}
            Suppose Assumptions~\ref{assum:Lipschitz} and \ref{assum:gamma} hold. The original system under the proposed MPC controller~\eqref{eq:mpc} is input-to-state stable with respect to the approximation error $\epsilon_k$.
    \end{rm}
\end{corollary}

The closed-loop nonlinear system is input-to-state stable with respect to the Koopman model approximation error; therefore, reducing this approximation error as much as possible is key to improving control performance.
This directly connects to the prediction error characterized in Lemma~\ref{lem:estimation_error}. As established in Lemma~\ref{lem:estimation_error}, the prediction error depends on the width of the DNN hidden layers, the system's variation rates $(\mu_x,\mu_u)$, and the Lipschitz property of the network $\mu_g$. Consequently, improving the tracking performance requires optimizing these factors.
For example, this can be achieved by increasing the number of nodes in the hidden layers, using a higher sampling frequency (i.e., a smaller discretization time step), and more frequently updating the lifted system matrices.

\section{Numerical simulations}\label{sec:simulation}
In this section, four numerical examples are presented to demonstrate the effectiveness and advantages of the proposed algorithm.\footnote{Code available at: {https://github.com/wulidede/Online-time-varying-deep-Koopman-learning.git}.}
The first is a numerical example adopted from \cite{zhangOnlineDynamicMode2019,haoDeepKoopmanLearning2024}.
The second is the Duffing oscillator \cite{khalil2002nonlinear}, which serves as a representative time-varying nonlinear system.
The third example is a 3-DOF robotic arm system from \cite{singh2025adaptive}, illustrating the application of the proposed method to trajectory tracking control.
Finally, the effectiveness of the proposed algorithm is further validated on a synthetic biological network system in \cite{han2022desko}.

\subsection{A simple time-varying nonlinear system}
Consider a simple time-varying nonlinear system,
\begin{equation}
    \begin{aligned}
        \begin{bmatrix}
            \dot{x}_1(t)\\
            \dot{x}_2(t)
        \end{bmatrix}=\begin{bmatrix}
            0 & 1+\gamma t\\
            -(1+\gamma t) & 0
        \end{bmatrix}\begin{bmatrix}
            \cos x_1(t)\\
            \cos x_2(t)
        \end{bmatrix},
    \end{aligned}
\end{equation}
where $\gamma$ is a constant parameter that controls the rate of change in the system.
Details about this system can be found in \cite{zhangOnlineDynamicMode2019,haoDeepKoopmanLearning2024}.

\textbf{Experimental parameter setting:} The experimental parameter settings are listed in Table~\ref{tab:sim_parameters}.
The continuous-time system is discretized using a zero-order hold and the Runge-Kutta method with $dt=0.1s$.
Three methods are used to predict the system's state: TVDMD \cite{zhangOnlineDynamicMode2019}, DKTV \cite{haoDeepKoopmanLearning2024}, and the method proposed in this paper.
The latter two methods use DNNs as observables with the same structure and training parameters.
The DNN $g(\cdot,\theta): \mathbb{R}^2 \rightarrow \mathbb{R}^6$ used as the lifting function has the architecture $[6,16,256,6]$, meaning that it consists of an input layer with 2 nodes, two hidden layers with 16 and 256 nodes, respectively, and an output layer with 6 nodes, employing ReLU as the activation function.

This simple example is intended to verify the effectiveness of the iterative update method proposed in Theorem~\ref{update_rule}, compared to existing methods. 
Therefore, conditions (\ref{eq:event_trigger}) and (\ref{eq:event_trigger2}) are not taken into account during each update in this simulation.
Both of this system and the Duffing oscillation system are converted to discrete systems using the Runge-Kutta method, with a discretization time interval of $dt=0.1$s.
The system's initial value is $x_0=[1,0]^{\top}$, for a simulation time of $t_{all}=30$s. 
DKTV is initialized using the first $t_{ini}=10$ seconds of data, while OTVDKL is initialized with data in the 10th second. 
The remaining 20 seconds of data are used for prediction to compare the performances of the three methods.
\begin{table}[htbp]
\caption{Summary of Simulation Parameters for Three Numerical Examples}
\label{tab:sim_parameters}
\centering
\begin{tabular}{c p{6cm}} 
\hline\hline
\textbf{Scenarios} & \multicolumn{1}{c}{\textbf{Parameters}} \\
\hline
\multirow{3}{*}{Section 5.1} 
& $\gamma=6, x_0=[1, 0]^\top$, $dt = 0.1s$, $t_{all}= 30s$; \\
& DNN: [2,16,256,6], ReLU, $t_{ini}= 10s$. \\
\hline
\multirow{4}{*}{Section 5.2} 
& $\alpha=1, \delta=0.2, \beta=0.5, \gamma=0.3, \omega=1.3$; \\
& $dt = 0.1s, x_0=[1, 0]^\top$; \\
& $w=30$, $b=10$, $\epsilon=0.15$; \\
& DNN: [2,16,256,6], ReLU, $t_{ini}= 3s$. \\
\hline
\multirow{5}{*}{Section 5.3} 
& $\forall i : m_i=0.6\text{kg},l_i=1\text{m},I_i=\text{diag}(0,m_il_i^2/12,m_il_i^2/12)\text{kgm}^2, dt = 0.01\text{s}$;\\
& $\tau \in [-30, 30]^3$, $w=50 $, $b= 5$; \\
& DNN: [6,30,30,16], ReLU; Horizon $H=15$,\\
&  $Q=\text{diag}(200I_6,{0}_{10\times10})$, $R=10I_3$; \\
\hline
\multirow{5}{*}{Section 5.4} 
& $\forall i : K_i = 1, a_i = 1.6,\gamma_i = 0.16, \beta_i = 0.16, c_i = 0.06, \omega=1, dt = 1s$;\\
& $x_0 \in [0, 5]^6, u \in [0, 5]^3$, $w=200 $, $b= 100$; \\
& DNN: [6,100,80,16], ReLU, $t_{ini}= 20$; \\
& $H=16$, $Q=\text{diag}(I_6,{0}_{10\times10})$, $R=0.1I_3$; \\
\hline\hline
\end{tabular}
\end{table}
\begin{figure}
    \begin{center}
    \includegraphics[height=10cm]{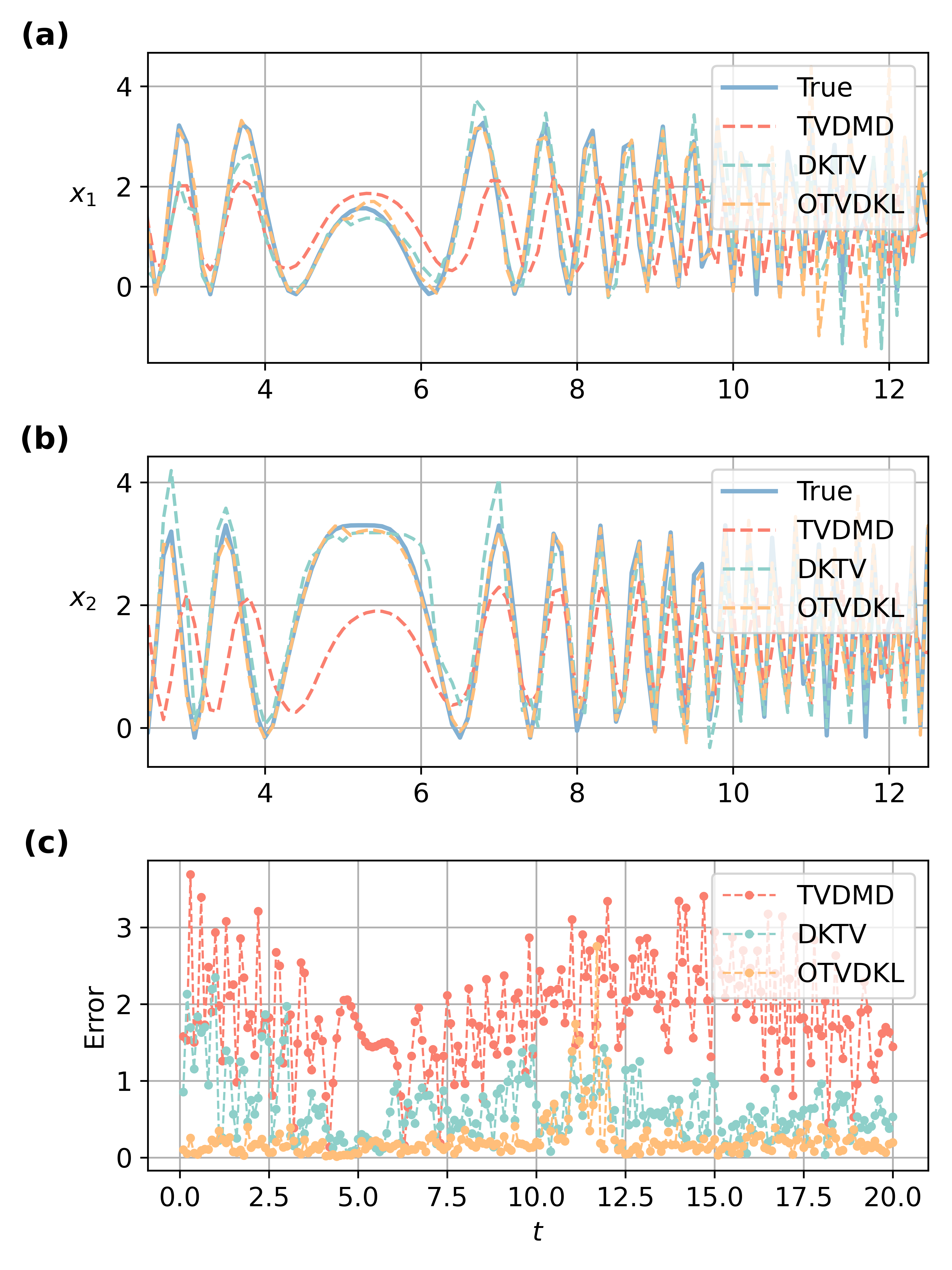}  
    \caption{Prediction results and errors of the three methods. Subfigures (a) (b) and (c) represent the trajectory predictions of \(x_1\), the trajectory predictions of \(x_2\), and the prediction error plots for the three methods, respectively.}  
    \label{fig:NTVS}     
    \end{center}     
\end{figure}

\textbf{Results and analysis:} The prediction results of the three methods are shown in Fig.~\ref{fig:NTVS}.
As can be seen from this figure, the predicted state of the proposed method exhibits a high degree of consistency with the state of the system, demonstrating a notable reduction in prediction error compared to the other two methods.
To provide a more quantitative comparison of the prediction errors, the mean absolute error (MAE) is used, which is defined as follows:
\begin{equation}
    MAE:=\frac{1}{N}\sum_{k=1}^{N}\|x(k)-x_{\text{pred}}(k)\|_2,
\end{equation}
where $x_{\text{pred}}(k)$ is the predicted state of the system at time step $k$.

The proposed online time-varying deep Koopman learning method is abbreviated as OTVDKL.
The MAE and RMSE of the predictions for the three methods are shown in Table \ref{NTVS_error}.
\begin{table}[!h]
    \renewcommand{\arraystretch}{1.2}
    \caption{The error evaluation of the three methods.}
    \label{NTVS_error}
    \centering
    \begin{tabular}{c c c c c}
        \hline
         & TVDMD \cite{zhangOnlineDynamicMode2019} & DKTV \cite{haoDeepKoopmanLearning2024} & OTVDKL \\
        \hline
        MAE & 1.865 & 0.6186 & 0.1846 \\
        RMSE & 1.980 & 0.7353 & 0.2245 \\
        \hline
    \end{tabular}
\end{table}
As evidenced by the table, the proposed method achieves a notable reduction in both MAE and RMSE, with 90.1\% and 88.7\% decrease, respectively, in comparison to TVDMD \cite{zhangOnlineDynamicMode2019}. 
In comparison to DKTV \cite{haoDeepKoopmanLearning2024}, the MAE and RMSE are reduced by 70.2\% and 69.4\%, respectively.
The significant reduction in prediction error demonstrates the effectiveness of the proposed iterative update method, which obtains the lifted system matrices by using only the most recent snapshots for computation.

\subsection{Duffing oscillator}
\begin{figure*}
    \begin{center}
    \includegraphics[scale=0.26]{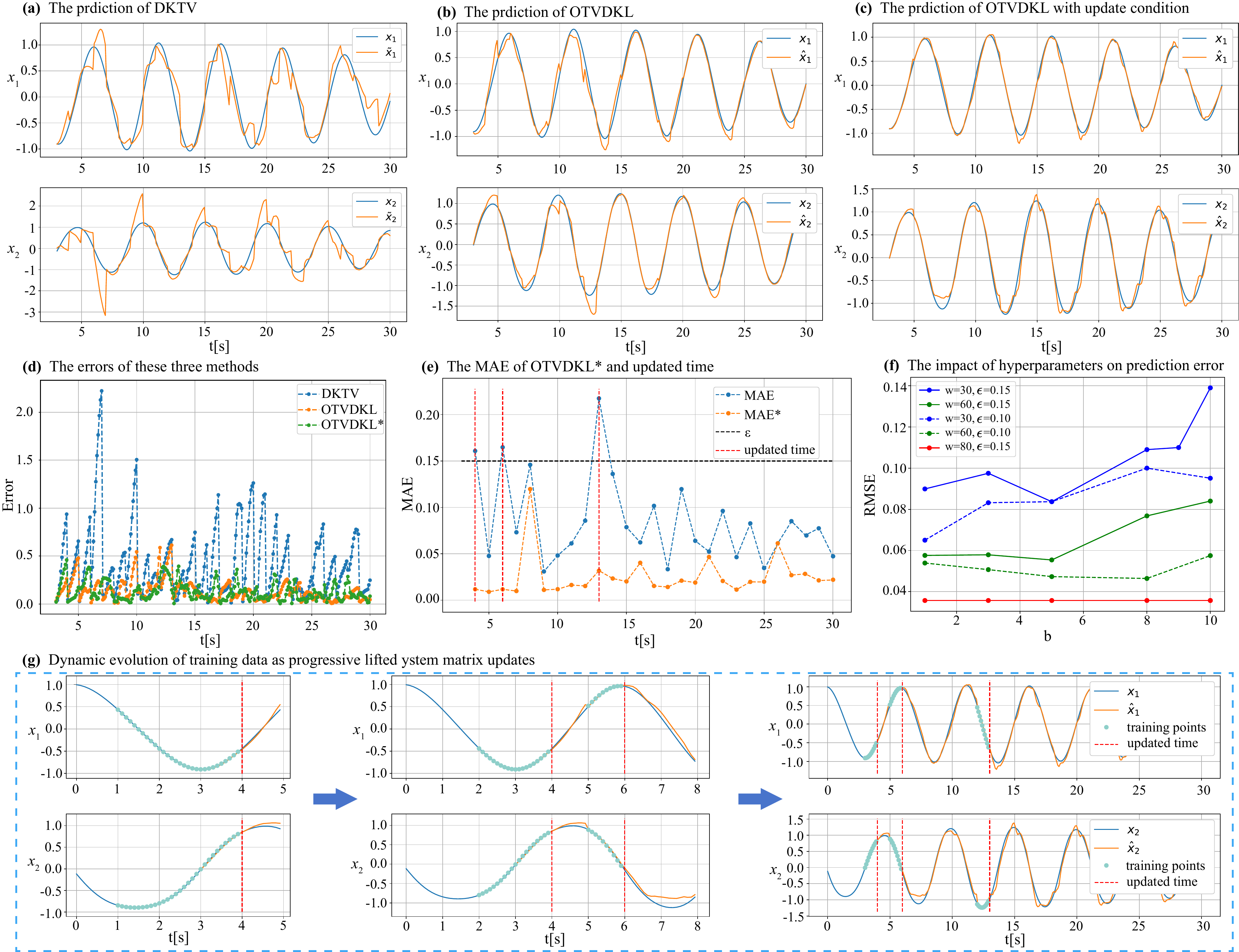}  
    \caption{{Prediction results and errors for the Duffing oscillator. (a) The prediction results of the DKTV. (b) The prediction results of the OTVDKL. 
    (c) The prediction results of the OTVDKL*. In these three subfigures, the blue curves represent the ground truth, while the orange curves correspond to the estimates produced by the respective methods.
    (d) The absolute error over time for DKTV (Blue), OTVDKL (Orange), and OTVDKL* (Green). 
    (e) The MAE of OTVDKL* at each evaluation step. Blue dots represent the MAE of the current model on new data; Orange dots represent the MAE after a potential model update. The vertical red dashed lines indicate the time instances where the update condition~\eqref{eq:event_trigger} was met, and the model was updated.
    (f) The impact of hyperparameters on the RMSE of prediction error.
    (g) Evolution of Training Data: Visualization of the iterative update process. The blue arrows indicate the sliding window mechanism. 
    The blue solid curve and the orange solid curve represent the true values and the predicted values, respectively, while the blue scatter points denote the data samples in $S_{\tau}^{\text{cur}}$. As time progresses, the dataset $\mathcal{S}_{\tau}^{\text{cur}}$ selectively shifts to incorporate new data while discarding outdated snapshots.}
}
    \label{fig:duffing}     
    \end{center}     
\end{figure*}
The proposed method is tested on a complex time-varying nonlinear system, the Duffing oscillator, which is described as follows:
\begin{equation}
    \begin{aligned}
        \begin{bmatrix}
            \dot{x}_1(t)\\
            \dot{x}_2(t)
        \end{bmatrix}= \begin{bmatrix}
            x_2(t)\\
            -\alpha x_1(t)-\delta x_2(t)-\beta x^3+\gamma \sin(\omega t)
        \end{bmatrix},
    \end{aligned}
\end{equation}
where the system parameters are shown in Table~\ref{tab:sim_parameters}.

\textbf{Experimental parameter setting:}
In this scenario, three methods are used for testing: DKTV, OTVDKL without considering (\ref{eq:event_trigger}) and (\ref{eq:event_trigger2}) (denoted as OTVDKL), and OTVDKL with update condition (\ref{eq:event_trigger}) and (\ref{eq:event_trigger2}) (denoted as OTVDKL*).
The introduction of OTVDKL vs. OTVDKL* is intended for an ablation study to evaluate the effect of the update mechanisms~\eqref{eq:event_trigger} and \eqref{eq:event_trigger2}.
The neural network architecture is \([2,16,256,6]\), where \(x_1\) and \(x_2\) are used as inputs, followed by two hidden layers with 16 and 256 neurons, respectively, and a six-dimensional high-dimensional state as the output.
The experimental parameter settings are listed in Table~\ref{tab:sim_parameters}.
The DNN training  process is implemented by choosing the optimizer Adam with learning rate = 1e-3 and weight decay rate = 1e-4.
The same training configuration is also applied in the subsequent two examples.
In this simulation, we choose a window width of $w = 30$, an update interval of $b = 10$, and $\epsilon = 0.15$ in~(18).
Note that using slightly larger values of \( w \) can improve numerical stability.
Adding a regularization term $\lambda I$ (e.g., $\lambda = 10^{-3}$) to the data matrix improves numerical stability during online updates.
Due to the stochastic nature of neural network training, we repeat each method ten times and report the mean and standard deviation of the tracking error.

\textbf{Results and analysis:} The prediction results and errors of the three methods are shown in Fig.~\ref{fig:duffing}.
The prediction results of the three methods are shown in Figs.~\ref{fig:duffing}(a), (b), and (c), respectively. 
The predictions given by DKTV show a notable difference from the true trajectory, which is attributed to the fact that it uses outdated data to compute the Koopman system matrix.
The OTVDKL, by using only the most recent data for computation, results in relatively small prediction errors. 
However, due to the lack of consideration of the update condition (\ref{eq:event_trigger}) and the indiscriminate shifting of the window for updates, it occasionally leads to larger errors, especially near the system's peak state.
The OTVDKL* achieves smaller prediction errors compared to the other two methods and provides better predictive performances at the system's peak states.
These conclusions are supported by the results shown in Fig.~\ref{fig:duffing}(d), which illustrates the prediction errors of the three methods.
\ref{fig:duffing}(g) shows the evolution of the dataset $\mathcal{S}_{\tau}^{\text{cur}}$ as the model is updated. 
After three updates, the dataset $\mathcal{S}_{\tau}^{\text{cur}}$ no longer changes, indicating that it has sufficiently well captured the system's dynamics.

Fig.~\ref{fig:duffing}(e) illustrates the MAE for predictions on the new snapshots in $\mathcal{S}_{\tau}^{\text{new}}$ at every evaluation update time, using the non-updated model and the updated model, corresponding to the right-hand and left-hand sides of (\ref{eq:event_trigger}), respectively.
At most time instants, the MAE of the non-updated model's predicted state is below the specified threshold $\epsilon$, indicating that the lifted system matrices do not need to be updated. This requires using the update condition described in (\ref{eq:event_trigger2}) to determine whether an update is necessary.
The prediction error of the updated model is usually smaller than that of the previous model. 
Nevertheless, there are instances when the prediction error of the updated model actually increases, which necessitates meeting the update condition described in (\ref{eq:event_trigger}).
By introducing the update conditions (\ref{eq:event_trigger}) and (\ref{eq:event_trigger2}), many unnecessary updates are avoided, significantly reducing the overall computational cost. 
The effects of the window width \(w\), the update interval \(b\), and the threshold \(\epsilon\) on the prediction RMSE are illustrated in Fig.~\ref{fig:duffing}(f).
It can be observed that, for the prediction task, a larger \(w\) leads to a smaller RMSE, which is likely due to the fact that more data are utilized. Moreover, decreasing \(b\) and the threshold \(\epsilon\) increases the update frequency, which in turn reduces the prediction error.
The resulting data for MAE, RMSE, and computational time of the three methods are summarized in Table \ref{duffing}.
\begin{table}[t]
    \centering
    \caption{Comparison of MAE, RMSE, and Computation Time for different methods (Mean $\pm$ Std).}
    \label{duffing}
    \begin{tabular}{lccc}
        \toprule
        Method & MAE & RMSE & Time (s) \\
        \midrule
        DKTV    & $0.554 \pm 0.151$ & $0.792 \pm 0.245$ & $4.42 $ \\
        OTVDKL  & $0.222 \pm 0.052$ & $0.303 \pm 0.091$ & $3.14 $ \\
        OTVDKL* & $0.158 \pm 0.031$ & $0.194 \pm 0.050$ & $0.89 $ \\
        \bottomrule
    \end{tabular}
\end{table}
Compared to DKTV, the prediction errors of OTVDKL with update conditions are reduced by 71.5\% in MAE and 75.5\% in RMSE.
Since OTVDKL with update conditions significantly decreases unnecessary updates, the computational time is greatly reduced. 
The significant reduction in both prediction error and computational time demonstrates the effectiveness of the algorithm proposed in this paper.
\subsection{Serial manipulator}
\begin{figure}
    \begin{center}
    \includegraphics[height=4cm]{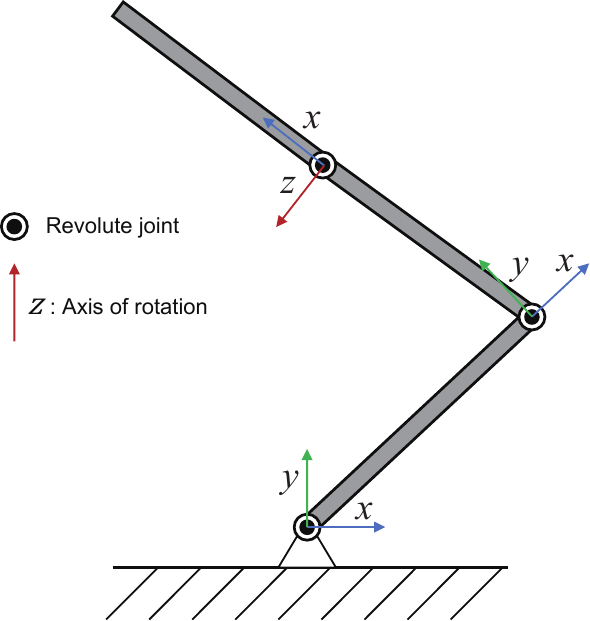}  
    \caption{Three-axis robotic manipulator in three-dimensional space. The rotation axis of the joint in the terminal link is orthogonal to the first two rotation axes.}  
    \label{fig:manipulator}     
    \end{center} 
\end{figure}

Next, we consider trajectory tracking of a three-axis robotic manipulator in three-dimensional space (as shown in Fig.~\ref{fig:manipulator}).
The system dynamics for a serial manipulator is given as
\begin{equation}\label{eq:serial}
M(\theta)\ddot{\theta} + C(\theta,\dot{\theta}) + G(\theta)= \tau + \tau_d,
\end{equation}
where $\theta \in \mathbb{R}^3$, $\tau \in \mathbb{R}^3$, and $\tau_d \in \mathbb{R}^3$ represent the joint angles, the torque inputs, and the external disturbance torque, respectively.
$M\in \mathbb{R}^{3\times 3}$, $C \in \mathbb{R}^{3\times 3}$, and $G \in \mathbb{R}^{3\times 1}$ represent the mass, the Coriolis, and the gravity matrix, respectively.
The mass, inertia, and length of the $i$th link are provided in Table~\ref{tab:sim_parameters}.
In~\eqref{eq:serial}, the external disturbance torque $\tau_d$ is modeled as a time-varying sinusoidal function, i.e., $\tau_d = [10\sin(0.5t), 10\sin(0.5t), 10\sin(0.5t)]^{\top}$.

\textbf{Experimental settings:} The inputs to the neural network consist of the three-axis joint angles and angular velocities.
The network architecture is configured as $[6, 30, 30, 16]$, yielding a 16-dimensional lifted linear system.
The initial neural network is trained using 10{,}000 randomly sampled snapshots over 500 epochs.
The reference trajectories for all three joints are given by $ \theta_{\mathrm{ref}} = 0.5 \sin(0.4\pi t + \pi)$.
During online operation, Gaussian noise $\mathcal{N}(0, 0.05^2)$ is added to the original state measurements and the parameters of the DNN are fixed.
The comparison methods include: the proposed algorithm (OTVDKL), DKO (i.e., MPC based on a fixed lifted linear system),
adaptive linear Koopman (ALK in~\cite{singh2025adaptive}), and simultaneous system identification and MPC (SSI in~\cite{zhou2025simultaneous}).
We use the efficient MPC solver ACADOS \cite{verschueren2022acados} to solve the MPC problem~\eqref{eq:mpc}.
Since the neural network performance is sensitive to initialization, each method is evaluated over 10 independent runs.
The remaining parameter settings are summarized in Table~\ref{tab:sim_parameters}.

\begin{figure}
    \begin{center}
    \includegraphics[height=9cm]{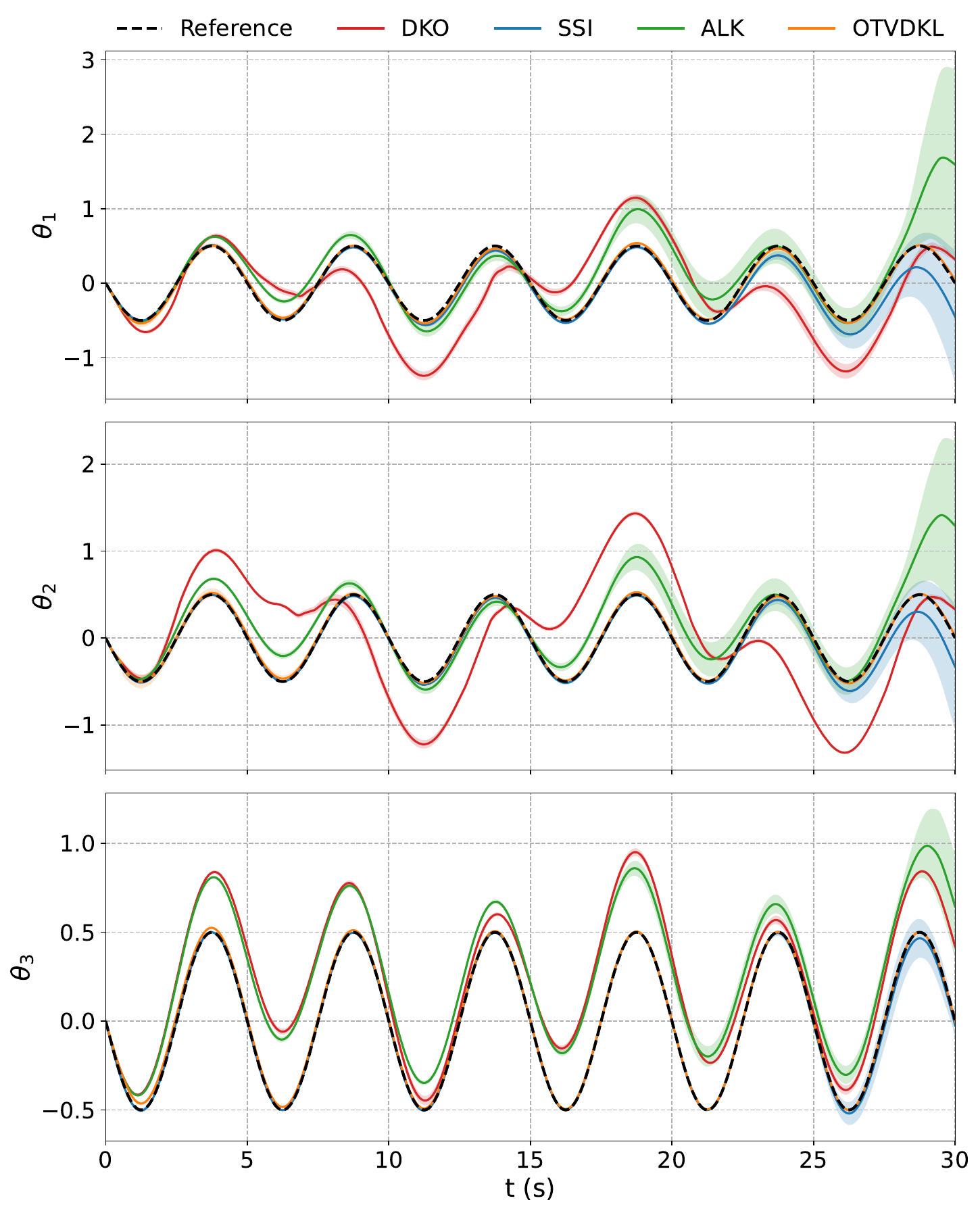}  
    \caption{Trajectory tracking performance of the serial manipulator (lines denote the mean performance, and shaded bands indicate the standard deviation for each method).}  
    \label{fig:tracking_results}     
    \end{center}     
\end{figure}
\textbf{Results and analysis:} Fig.~\ref{fig:tracking_results} illustrates the tracking performance of the three angular references achieved by several methods.
We plot the mean trajectories together with the standard-deviation error bands.
As shown in Fig.~\ref{fig:tracking_results}, the DKO method exhibits the poorest tracking performance, with relatively large tracking errors, due to its use of a fixed high-dimensional linear model.
In contrast, the SSI and ALK methods incorporate online update mechanisms and therefore significantly reduce the tracking error.
Benefiting from the robust controller design, the proposed method (OTVDKL) achieves the smallest tracking error among all methods.
The mean and standard deviation of the tracking errors are reported in Table~\ref{tab:serial_results}.
\begin{table}[htbp]
    \centering
    \caption{Performance comparison of different methods (Mean $\pm$ Std).}
    \label{tab:serial_results}
    \begin{tabular}{lcc}
        \toprule
        Method & RMSE & MAE \\
        \midrule
        DKO    & $0.4737 \pm 0.0112$ & $0.3906 \pm 0.0093$ \\
        SSI    & $0.0942 \pm 0.0813$ & $0.0665 \pm 0.0391$ \\
        ALK    & $0.2358 \pm 0.1241$ & $0.1661 \pm 0.0626$ \\
        OTVDKL & ${0.0688 \pm 0.0069}$ & ${0.0419 \pm 0.0043}$ \\
        \bottomrule
    \end{tabular}
\end{table}
In addition, we examine the impact of the hyperparameters \(w\) and \(b\) on the tracking error. The RMSE achieved by the proposed method under different settings of \(w\) and \(b\) is shown in Fig.~7.
Unlike the previous example, where \(w\) had a different effect on the prediction error, a larger \(w\) leads to a larger tracking error in this case.
This may be attributed to the fact that, for this system, a larger \(w\) incorporates more outdated information, which is detrimental to tracking control.
Similarly, a smaller \(b\), corresponding to more frequent updates, results in a smaller tracking error.
\begin{figure}
    \begin{center}
    \includegraphics[height=4.5cm]{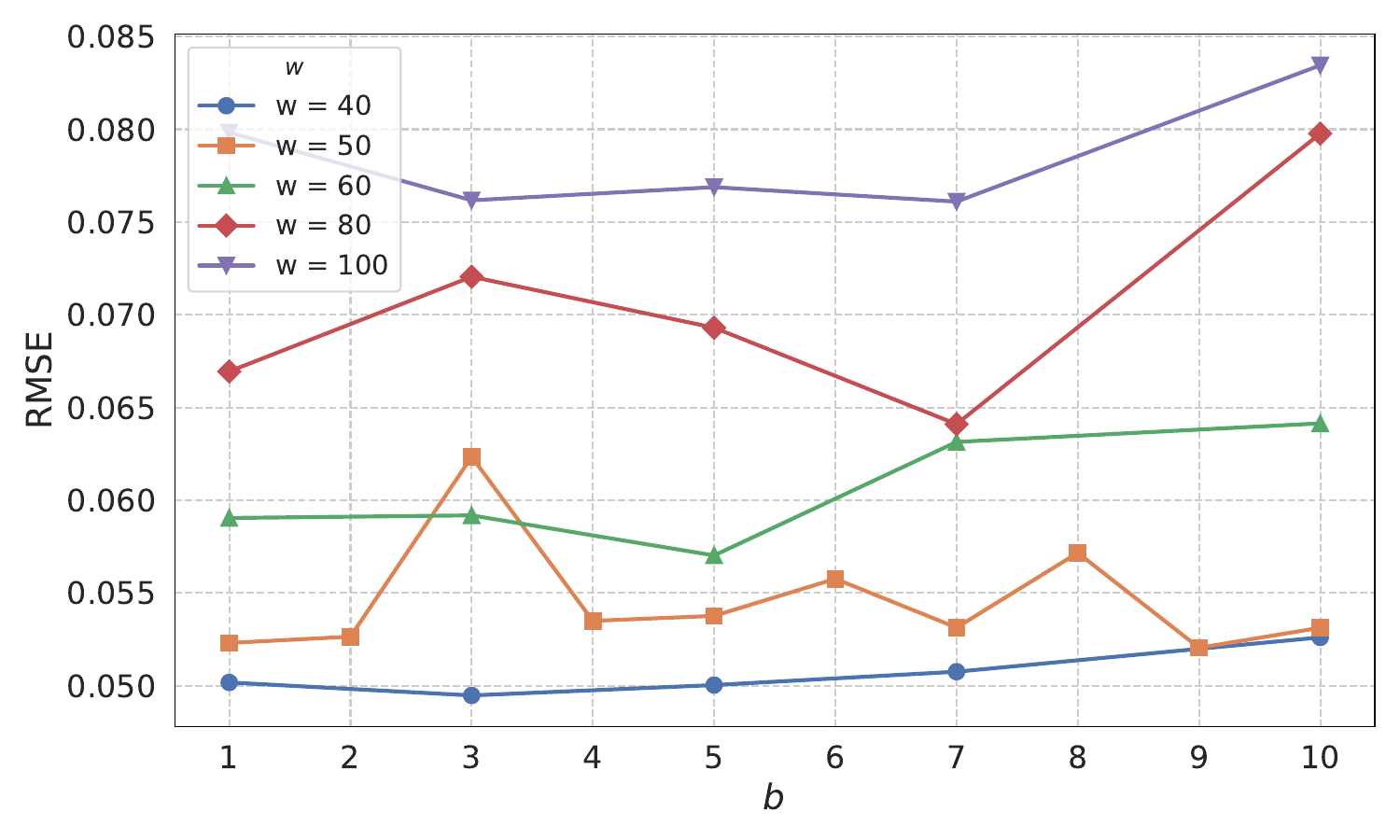}  
    \caption{The impact of \(w\) and \(b\) on the tracking error.}  
    \label{fig:tracking_wb}     
    \end{center}     
\end{figure}

\subsection{ Synthetic biological network system}\label{sec:bio_network}
In this example, we use a classic dynamical system  in systems and synthetic biology to demonstrate the effectiveness of the proposed algorithm in predicting and controlling nonlinear time-varying systems.

\begin{figure}
    \begin{center}
    \includegraphics[height=6.3cm]{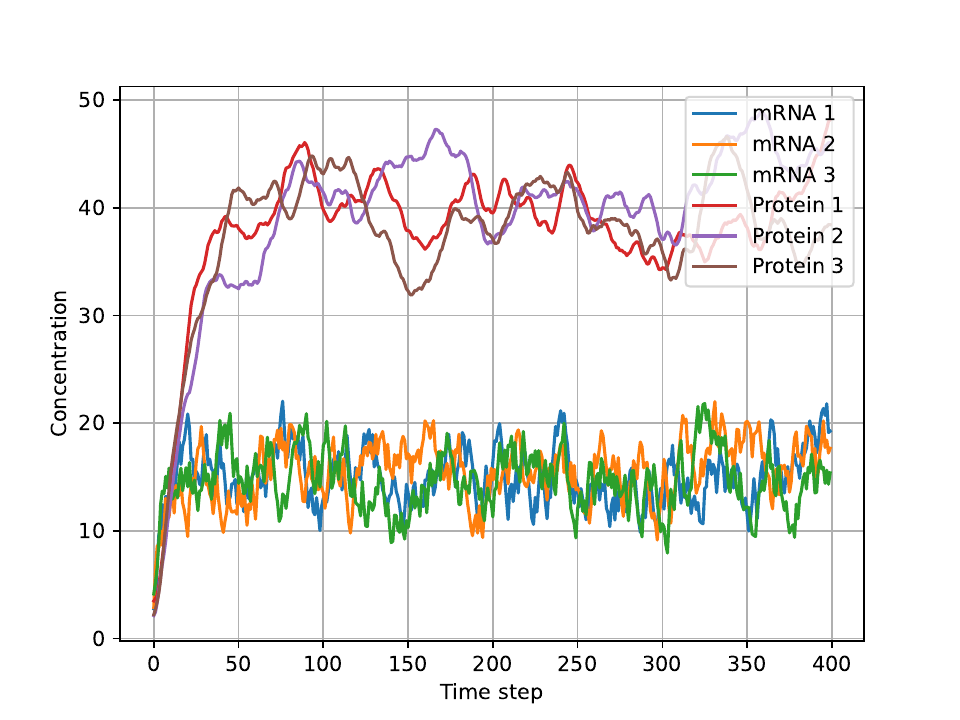}  
    \caption{The trajectory of the synthetic biological network system.}  
    \label{fig:bio_network}     
    \end{center}     
\end{figure}

The gene regulatory networks (GRN) is a synthetic regulatory network comprising three genes, whose mRNA and protein dynamics exhibit oscillatory behavior~\cite{elowitz2000synthetic}. 
Here, $x_1,x_2,x_3 (\text{or } x_4,x_5,x_6)$ represents the concentrations of the mRNA transcripts (or protein) of genes 1,2 and 3, respectively.
A detailed physical description of the system can be found in~\cite{han2022desko}.
A discrete-time model capturing both transcription and translation processes is described by the following equations:
\vspace{-0.3cm}
\begin{small}
\begin{equation}
    \begin{aligned}
        x_1(t+1) =& x_1(t) + {d}t\bigg[ -\gamma_1x_1(t) + \frac{a_1}{K_1\sin(\omega t) + x_6^2(t)} + u_1 \bigg], \\
        x_2(t+1) = &x_2(t) + {d}t\bigg[ -\gamma_2x_2(t) + \frac{a_2}{K_2\sin(\omega t) + x_4^2(t)} + u_2 \bigg], \\
        x_3(t+1) = & x_3(t) + {d}t\bigg[ -\gamma_3x_3(t) + \frac{a_3}{K_3\sin(\omega t) + x_5^2(t)} + u_3 \bigg], \\
        x_4(t+1) =& x_4(t) + {d}t[-c_1x_4(t) + \beta_1x_1(t)], \\
        x_5(t+1) =& x_5(t) + {d}t[-c_2x_5(t) + \beta_2x_2(t)], \\
        x_6(t+1) =& x_6(t) + {d}t[-c_3x_6(t) + \beta_3x_3(t)].
    \end{aligned}
\end{equation}
\end{small}
The term \(\sin(\omega t)\) is added as a time-varying component to represent the periodic variation of parameters.
The initial values of the states are randomly selected from the interval \([0, 5]^6\), and the control inputs \(u_i\) are randomly chosen from \([0, 5]^3\).
A true trajectory of 400 steps is generated, as shown in Fig. \ref{fig:bio_network}.
It can be observed that the state of this nonlinear time-varying system changes drastically.

\subsubsection{Prediction performance}
Three methods are compared in terms of prediction performance: the proposed algorithm (OTVDKL), DKTV, and the deep Koopman operator (DKO). The DKO  uses the same neural network as the lifting function but does not perform any online updates.
The simulation parameters are listed in Table~\ref{tab:sim_parameters}.
We generate $10$ test trajectories, each consisting of 4000 steps, by randomly sampling the system's initial states and control inputs.
Fig. \ref{fig:bio_network} shows one of the ten randomly generated trajectories.
Each trajectory is evaluated through \textit{four} repeated experiments.

\begin{figure}
    \begin{center}
    \includegraphics[height=6.3cm]{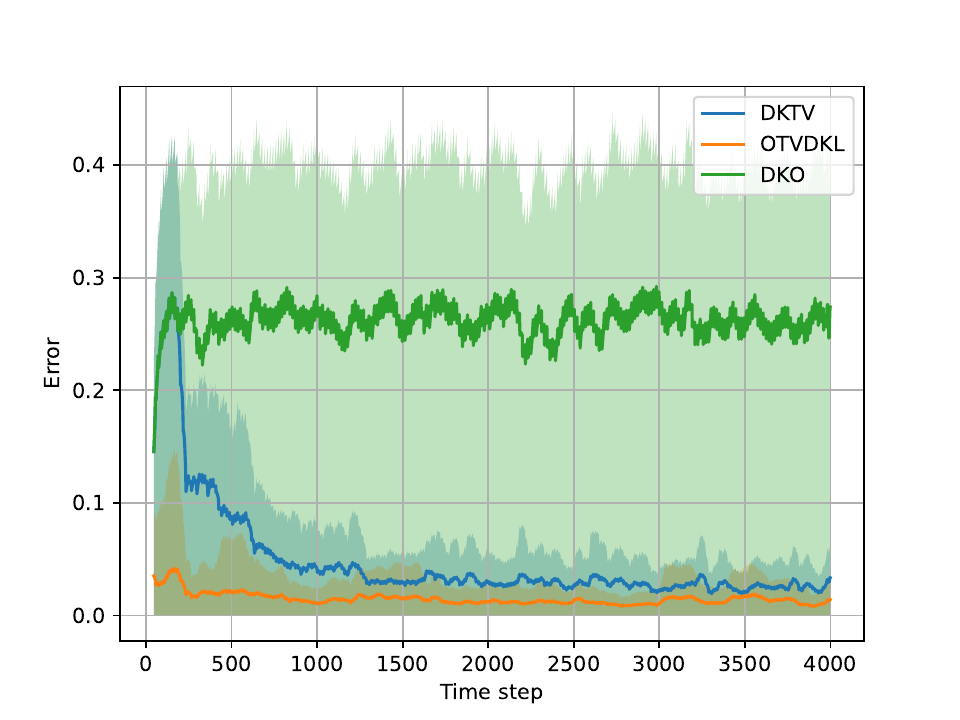}  
    \caption{The mean and standard deviation of the norm of the prediction error.
    The curves represent the mean prediction errors of DKO (green), DKTV (blue), and the proposed OTVDKL (orange), respectively.
    The shaded areas indicate the standard deviation calculated across 40 independent experiments at each time step.    } 
    \label{fig:error}     
    \end{center}     
\end{figure}
All three methods are evaluated on the same ten trajectories with four repeated experiments each.
The mean and standard deviation of the prediction errors over the 40 experiments are shown in Fig. \ref{fig:error}.
As can be seen from Fig. \ref{fig:error}, the deep Koopman operator achieves good prediction performance at the beginning.
However, as time progresses, the prediction error increases rapidly and then remains at a high level.
This indicates that, for nonlinear time-varying systems, using a Koopman operator trained offline can only yield accurate predictions within the range of the training data.
The mean prediction errors of the deep Koopman operator, DKTV, and our proposed method are 0.262, 0.0485, and 0.0150, respectively.
Compared to the DKTV proposed in \cite{haoDeepKoopmanLearning2024}, our proposed OTVDKL algorithm achieves smaller prediction errors, demonstrating a significant advantage in state prediction for unknown nonlinear time-varying controlled systems.
\subsubsection{Control performance}
This part verifies the effectiveness of the proposed control algorithm.
The control objective is to make the concentration of protein~1, i.e., \(x_4\) in the state vector, track a predefined trajectory, which is set as a sine wave.
DKTV~\cite{haoDeepKoopmanLearning2024} and DKO are designed solely for prediction and do not provide controller designs.
Therefore, we attach an MPC controller to them for comparative evaluation. The remaining parameter settings can be found in Table~\ref{tab:sim_parameters}.
\begin{figure}
    \begin{center}
    \includegraphics[height=6.3cm]{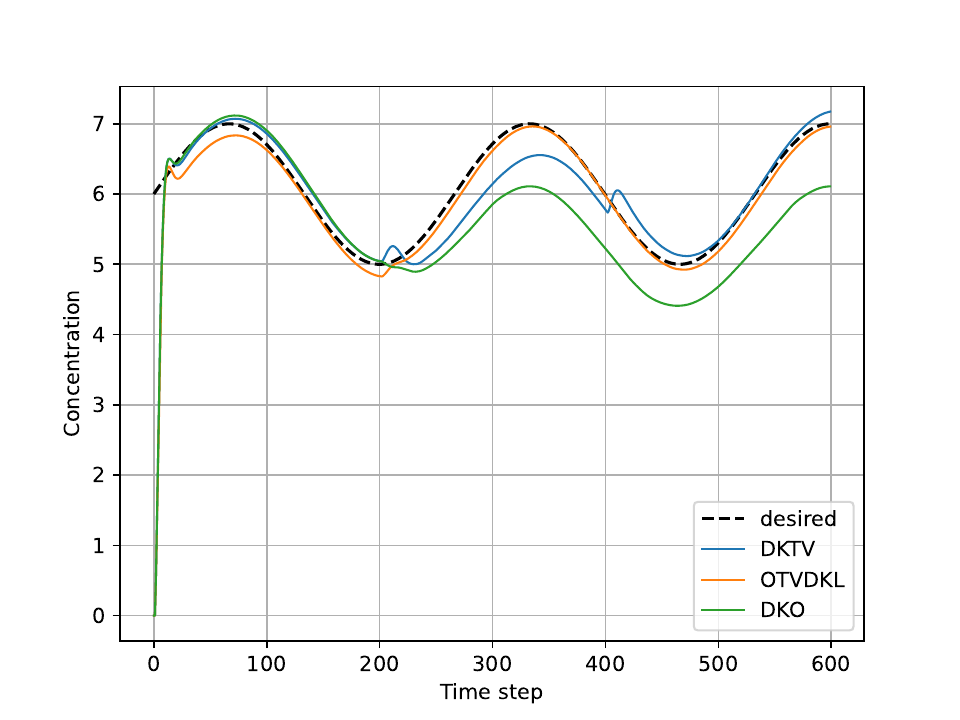}  
    \caption{ The tracking performance of the control task. The black dashed line represents the desired trajectory.} 
    \label{fig:tracking}     
    \end{center}     
\end{figure}

The tracking performance of the control task is shown in Fig. \ref{fig:tracking}.
It can be observed that the MPC controller based on the Koopman operator trained with offline collected data consistently exhibits a large tracking error with respect to the reference trajectory.
{Without the online update (as seen in DKO), the MPC controller alone cannot handle the system variations effectively.
}The MPC controller based on DKTV also shows a noticeable error relative to the reference trajectory.
The tracking errors of the deep Koopman operator, DKTV, and our proposed method are 0.623, 0.269, and 0.106, respectively.
The closed-loop system resulting from our proposed OTVDKL method combined with controller~\eqref{eq:mpc} demonstrates excellent tracking performance, confirming that the proposed algorithm effectively approximates unknown nonlinear time-varying systems while enabling a well-performed controller with guaranteed stability.

\section{Conclusion}\label{sec:conclusion}
This paper proposes an online deep Koopman learning method for predicting and controlling unknown nonlinear time-varying systems.
Compared to the static approaches~\cite{luschDeepLearningUniversal2018} or the accumulative schemes~\cite{haoDeepKoopmanLearning2024,zhangOnlineDynamicMode2019}, the proposed method incorporates a sliding-window strategy to explicitly discard outdated data and a selective update mechanism to reduce computational complexity.
Based on the learned model, a controller with provable stability is designed, ensuring that the closed-loop system is input-to-state stable with respect to the Koopman approximation error.
Simulation results demonstrate the superiority of the proposed algorithm in state prediction and its effectiveness in tracking control.

\appendix
\section{Proof of Proposition~\ref{update_rule}}\label{apdx:A}    
\begin{pf}
Under Assumption~\ref{assum:full_rank}, the explicit solution obtained by minimizing the loss function (\ref{eq:loss_compact}) is given in (\ref{eq:AB}). 
Expanding this yields
\begin{equation}
    \begin{bmatrix}
        A_{\tau} & B_{\tau}\\
    \end{bmatrix}=H_{\tau}\begin{bmatrix}
        G_{\tau} \\
        U_{\tau}
    \end{bmatrix}^{\top}\left( \begin{bmatrix}
        G_{\tau} \\
        U_{\tau}
    \end{bmatrix}\begin{bmatrix}
        G_{\tau} \\
        U_{\tau}
    \end{bmatrix}^{\top}\right)^{-1}.
\end{equation}
Set
\begin{equation}
    Q_{\tau}=H_{\tau}\begin{bmatrix}
        G_{\tau} \\
        U_{\tau}
    \end{bmatrix}^{\top},\quad P_{\tau}=\left(\begin{bmatrix}
        G_{\tau} \\
        U_{\tau}
    \end{bmatrix}\begin{bmatrix}
        G_{\tau} \\
        U_{\tau}
    \end{bmatrix}^{\top}\right)^{-1}.
\end{equation}
Then
\begin{equation}
    \begin{bmatrix}
        A_{\tau} & B_{\tau}\\
    \end{bmatrix}
    =Q_{\tau} P_{\tau}.
\end{equation}
If Assumption~\ref{assum:full_rank} holds at the update instant $\tau+1$, then matrices $A_{\tau+1}$, $B_{\tau+1}$, $Q_{\tau+1} \text{ and }P_{\tau+1}$ have the same relation as described above and $P_{\tau+1}$ is invertible.
Therefore,
\begin{equation}
    \begin{aligned}
        Q_{\tau+1}=Q_{\tau}+W_{\tau}EZ_{\tau}^{\top},\\
        P_{\tau+1}^{-1}=P_{\tau}^{-1}+Z_{\tau}EZ_{\tau}^{\top},
    \end{aligned}
\end{equation}
where $E,Z_{\tau}\text{ and  }W_{\tau}$ are defined in Proposition~\ref{update_rule}.
According to the Sherman–Morrison–Woodbury formula \cite{max1950inverting}, one has
{\begin{equation}
    \begin{aligned}
        (A+UCV)^{-1}=&\\
        A^{-1}-A^{-1}&U{(C^{-1}+VA^{-1}U)}^{-1}VA^{-1},\\
    \end{aligned}
\end{equation}}
If $E+Z_{\tau}^{\top}P_{\tau}Z_{\tau}$ is invertible, then
\begin{equation}
    \begin{aligned}
        P_{\tau+1}&=(P_{\tau}^{-1}+Z_{\tau}EZ_{\tau}^{\top})^{-1}\\
        &=P_{\tau}-P_{\tau}Z_{\tau}(E+Z_{\tau}^{\top}P_{\tau}Z_{\tau})^{-1}Z_{\tau}^{\top}P_{\tau}.\\
    \end{aligned}
\end{equation}
Notably, the above expression uses the relation $E^{-1}=E$. Furthermore, one has
\begin{equation}
    \begin{aligned}
        &\begin{bmatrix}
            A_{\tau+1} & B_{\tau+1}\\
        \end{bmatrix}\\
        =&Q_{\tau+1}P_{\tau+1}\\
        =&(Q_{\tau}+W_{\tau}EZ_{\tau}^{\top})(P_{\tau}-P_{\tau}Z_{\tau}\Gamma_{\tau}Z_{\tau}^{\top}P_{\tau})\\
        =&\begin{bmatrix}
            A_{\tau} & B_{\tau}\\
        \end{bmatrix}\left(I-Z_{\tau}\Gamma_{\tau}Z_{\tau}^{\top}P_{\tau}\right)+W_{\tau}EZ_{\tau}^{\top}P_{\tau}\\
        &-W_{\tau}EZ_{\tau}^{\top}P_{\tau}Z_{\tau}\Gamma_{\tau}Z_{\tau}^{\top}P_{\tau}\\
        =&\begin{bmatrix}
            A_{\tau} & B_{\tau}\\
        \end{bmatrix}\left(I-Z_{\tau}\Gamma_{\tau}Z_{\tau}^{\top}P_{\tau}\right)+\\
        &W_{\tau}E\left(\Gamma_{\tau}^{-1}-Z_{\tau}^{\top}P_{\tau}Z_{\tau}\right)\Gamma_{\tau}Z_{\tau}^{\top}P_{\tau}\\
        =&\begin{bmatrix}
            A_{\tau} & B_{\tau}\\
        \end{bmatrix}+(W_{\tau}-\begin{bmatrix}
            A_{\tau} & B_{\tau}
            \end{bmatrix}Z_{\tau})\Gamma_{\tau}Z_{\tau}^{\top}P_{\tau},\\
    \end{aligned}
\end{equation}
where $\Gamma_{\tau}=(E+Z_{\tau}^{\top}P_{\tau}Z_{\tau})^{-1}$.
If $E+W_{\tau}^{\top}(H_{\tau}H_{\tau}^{\top})^{-1}W_{\tau}$ is invertible, therefore one can derive the iterative update method for $C_{\tau+1}$ analogically.
\end{pf}

\section{Proof of Lemma~\ref{update_condition}}\label{apdx:B}         
\textbf{Schur complement formula~\cite{horn2012matrix}:}
{Consider a block matrix $\mathbf{M} = \begin{bmatrix} A & B \\ C & D \end{bmatrix}$, where $M$, $A$, and $D$ are all square matrices. The determinant of $\mathbf{M}$ is given by:
\begin{equation}
    \det(\mathbf{M}) = 
    \begin{cases} 
        \det(A)  \det(D - C A^{-1} B) & \text{if } A \text{ is invertible}, \\
        \det(D)  \det(A - B D^{-1} C) & \text{if } D \text{ is invertible}.
    \end{cases}
\end{equation}}
\begin{pf}
    Assumption~\ref{assum:full_rank} holding at the $\tau$-th update implies that $\begin{bmatrix}
            G_{\tau} \\
            U_{\tau}
            \end{bmatrix}\begin{bmatrix}
                G_{\tau} \\
                U_{\tau}
                \end{bmatrix}^{\top}$ is invertible. 
    Using the properties of the Schur complement, if $P_{\tau}^{-1}$ is invertible, then one has
    \begin{equation}\label{eq:schur}
        \begin{aligned}\det\left(
            \begin{bmatrix}
                P_{\tau}^{-1} & Z_{\tau}\\
                Z_{\tau}^{\top} & -E
            \end{bmatrix}\right)&=\det\left(P_{\tau}^{-1}\right)\det\left(-E-Z_{\tau}^{\top}P_{\tau}Z_{\tau}\right)\\
            &=\det(-E)\det\left(P_{\tau}^{-1}+Z_{\tau}EZ{\tau}^{\top}\right).
        \end{aligned}
    \end{equation}
    If $E+Z_{\tau}^{\top}P_{\tau}Z_{\tau}$ is invertible, then the left-hand side of the above equation (\ref{eq:schur}) is non-zero, which implies that $P_{\tau+1}^{-1}=P_{\tau}^{-1}+Z_{\tau}EZ_{\tau}^{\top}$ is invertible, i.e., $\begin{bmatrix}
        G_{\tau+1}\\
        U_{\tau+1}
    \end{bmatrix}\begin{bmatrix}
        G_{\tau+1}\\
        U_{\tau+1}
    \end{bmatrix}^{\top}$ is full rank.
    Thus, $\begin{bmatrix}
        G_{\tau+1}\\
        U_{\tau+1}
    \end{bmatrix}$ is of full row rank and Assumption~\ref{assum:full_rank} holds at the $(\tau+1)$-th update.
    
    Conversely, if $\begin{bmatrix}
        G_{\tau+1}\\
        U_{\tau+1}
    \end{bmatrix}$ is of full row rank at the $(\tau+1)$-th update, i.e., $P_{\tau}^{-1}+Z_{\tau}EZ_{\tau}^{\top}$ is invertible, then the right-hand side of the equation (\ref{eq:schur}) is non-zero, which implies that $E+Z_{\tau}^{\top}P_{\tau}Z_{\tau}$ is invertible.
    
    Similarly, it can be proved that $E+W_{\tau}^{\top}\left(H_{\tau}H_{\tau}^{\top}\right)^{-1}W_{\tau}$ being invertible is equivalent to $H_{\tau+1}$ being of full rank. 
    Combining two parts completes the proof of Lemma~\ref{update_condition}.
\end{pf}

\section{Proof of Lemma~\ref{lem:estimation_error}}\label{apdx:error}
\begin{pf}
    The proof of this lemma follows the proof of Theorem~1 in~\cite{haoDeepKoopmanLearning2024}.
    Recall that given the latest $\mathcal{S}_{\tau}^{\text{cur}}$ with $\{x_{k-1},u_{k-1}\}$ its latest data point (i.e., $x_{k-1}=x_{k_{\tau}+w-1},u_{k-1}=u_{k_{\tau}+w-1}$), $x_k$ denotes  the state of original system evolving from $x_{k-1}$ and $u_{k-1}$.
    Let $\hat{x}$ denote the estimated state by the proposed method. Then, the estimation error can be expressed as
    \begin{equation}\label{eq:error_decomp}
        \|e_k\|=\|x_k-\hat{x}_k+x_{k-1}-x_{k-1}\|.
    \end{equation}
    For any $\bar{x},\bar{u}\in\mathcal{S}_{\tau}^{\text{cur}}$ (note that here $x_{k-1} = \bar{x}_{k-1}, u_{k-1} = \bar{u}_{k-1}$ since $\{x_{k-1}, u_{k-1}\}$ is the observed state-input data point), we introduce $\epsilon_k$ as the local estimation error induced by the system approximation in the second term of loss~(\ref{eq:loss}) given by
    \begin{equation}
    \epsilon _ { k } = g ( \bar { x } _ { k } , \theta _ { \tau } ) - ( A _ { \tau } g ( \bar { x } _ { k - 1 } , \theta _ { \tau } ) + B _ { \tau } \bar { u } _ { k - 1 } ) .        
    \end{equation}
    Similarly, let $\bar{\epsilon}_k$ denote the reconstruction error induced by the lifting function in the first term of loss~(\ref{eq:loss}) by
    \begin{equation}
        \bar{\epsilon}_k = \bar{x}_k-C_{\tau}g(\bar{x}_k,\theta_{\tau}),
    \end{equation}
    which leads to 
    \begin{equation}\label{eq:C4}
        \bar{x}_{k-1}=C_\tau(A_\tau g(\bar{x}_{k-2},\theta_\tau)+B_\tau\bar{u}_{k-2})+C_\tau\epsilon_{k-1}+\bar{\epsilon}_{k-1}.
    \end{equation}
    By substituting $\hat{x}_k=C_\tau(A_\tau g(\hat{x}_{k-1},\theta_\tau)+B_\tau\hat{u}_{k-1})$ and equation (\ref{eq:C4}) into~\ref{eq:error_decomp}, we have
    \begin{equation}\label{eq:C5}
        \begin{aligned}\| e_{k}\|=&\| C_\tau A_\tau(g(\bar{x}_{k-2},\theta_\tau)-g(\bar{x}_{k-1},\theta_\tau))+C_\tau B_\tau\\&(\bar{u}_{k-2}-\bar{u}_{k-1})+C_{\tau}\epsilon_{k-1}+x_{k}-\bar{x}_{k-1}+\bar{\epsilon}_{k-1}\|,\\
            \overset{(a)}{\leq}&\parallel C_{\tau}A_{\tau}(g(\bar{x}_{k-2},\theta_{\tau})-g(\bar{x}_{k-1},\theta_{\tau}))\parallel+\\
            &\parallel C_{\tau}B_{\tau}(\bar{u}_{k-2}-\bar{u}_{k-1})\parallel+\parallel C_{\tau}\epsilon_{k-1}\parallel+\\
            &\parallel x_{k}-\bar{x}_{k-1}\parallel+\parallel\bar{\epsilon}_{k-1}\parallel,\\
            \overset{(b)}{\leq}&\parallel C_\tau A_\tau\parallel\parallel g(\bar{x}_{k-2},\theta_\tau)-g(\bar{x}_{k-1},\theta_\tau)\parallel+\\
            &\parallel C_\tau B_\tau\parallel\parallel\bar{u}_{k-2}-\bar{u}_{k-1}\parallel+\parallel C_\tau\parallel\parallel\epsilon_{k-1}\parallel+\\&\parallel x_k-\bar{x}_{k-1}\parallel+\parallel\bar{\epsilon}_{k-1}\parallel,\end{aligned}
    \end{equation}
    where (a) follows the triangle inequality and (b) is derived by subordinance and submultiplicativity.
    For the first two terms in the error expansion~\eqref{eq:C5}, Assumptions~\ref{assum:time-varying} and~\ref{assum:Lipschitz} yield
    \begin{equation}
        \| C_\tau A_\tau\|\| g(\bar{x}_{k-2},\theta_\tau)-g(\bar{x}_{k-1},\theta_\tau)\|\leq\|C_{\tau}A_{\tau}\|\mu_g\mu_x,
    \end{equation}and
    \begin{equation}
        \| C_\tau B_\tau\|\|\bar{u}_{k-2}-\bar{u}_{k-1}\|\leq\|C_{\tau}B_{\tau}\|\mu_u.
    \end{equation}
    For writing convenience, we denote $L_a:=\|C_{\tau}A_{\tau}\|\mu_g\mu_x+\|C_{\tau}B_{\tau}\|\mu_u$.
    As for the third term in the error expansion~\eqref{eq:C5}, following an argument similar to the proof in~\cite{haoDeepKoopmanLearning2024}, we derive the error bound of \( \|\varepsilon_k\| \) by backpropagating it from \( k \) to \( k_0 = 0 \).
    We define the global approximation error as
    \begin{equation}\label{eq:C8}
        \mathcal{E}_k=g(\bar{x}_k,\theta_\tau)-\prod_{i=0}^kA_ig(\bar{x}_0,\theta_0)-\sum_{j=0}^{k-1}(\prod_{l=0}^{k-j-1}A_l)B_ju_j.
    \end{equation}
    The related proof is inspired by the global accumulation rule of \( e_k \) proposed in \cite{mamakoukas2023learning}, given by $\mathcal{E}_k=\sum_{i=0}^{k-1}A^i\epsilon_{k-i}.$
    To achieve the error bound $\mathcal{E}_k$ in~\eqref{eq:C8}, it is necessary to replace $A$ with $A_{\tau}$ (note that the batch index $\tau$ is slower than the data point index $k$).
    Then \( \mathcal{E}_k \) is obtained through a recursive procedure, as detailed in \cite{haoDeepKoopmanLearning2024}.
    Here, we directly use the resulting expression:\\
    (1) when $\tau=1,k\in[k_1,k_1+w]$,
    \begin{equation}\label{eq:C9}
        \mathcal{E}_k=\sum_{i=0}^{k-k_1-1}A_1^i\epsilon_{k-i}+A_1^{k-k_1-1}\sum_{j=1}^{w}A_0^j\epsilon_{k_1+1-j},
    \end{equation}
    (2) when $\tau>1,k\in[k_\tau,k_\tau+w]$,
    \begin{equation}\label{eq:C10}
        \begin{aligned}
            \mathcal{E}_{k}&=\sum_{i=0}^{k-k_\tau-1}A_\tau^i\epsilon_{k-i}+A_\tau^{k-k_\tau-1}\Bigg(\sum_{j=1}^{w}A_{\tau-1}^j\epsilon_{k_\tau+1-j}+\\
            &\Big(\sum_{l=1}^{\tau-1}\sum_{m=1}^{w}(\prod_{n=\tau-1}^lA_n^{w})A_{l-1}^m\Big)\epsilon_{k_l+1-m}\Big),\end{aligned}
    \end{equation}
    (3) when $\tau+1,k\in [k_{\tau+1},k_{\tau+1}+w]$,
    \begin{equation}\label{eq:C11}
        \begin{aligned}\mathcal{E}_{k}&=\sum_{i=0}^{k-k_{\tau+1}-1}A_{\tau+1}^i\epsilon_{k-i}+A_{\tau+1}^{k-k_{\tau+1}-1}\biggl(\sum_{j=1}^{w}A_\tau^j\epsilon_{k_{\tau+1}+1-j}\\&+\Big(\sum_{l=1}^\tau\sum_{m=1}^{w}(\prod_{n=\tau}^lA_n^{w})A_{l-1}^m\Big)\epsilon_{k_l+1-m}\Big).\end{aligned}
    \end{equation}
    According to the above results~\eqref{eq:C9}-\eqref{eq:C11}, the third term of the error expansion~\eqref{eq:C5} can be bounded as
    \begin{equation}\label{eq:C12}
        \begin{aligned}\parallel C_{\tau}\epsilon_{k-1}\parallel\leq\parallel C_{\tau}&\bigg(\sum_{i=0}^{{k-k_{\tau}-2}}A_{\tau}^{i}+A_{\tau}^{{k-k_{\tau}-2}}\bigg(\sum_{j=1}^{{w}}A_{\tau-1}^{j}\\
            &+\sum_{l=1}^{{\tau-1}}\sum_{m=1}^{{w}}(\prod_{n=\tau-1}^{l}A_{n}^{{w}})A_{l-1}^{m}\bigg)\bigg)\parallel L_{1},\end{aligned}
    \end{equation}
    where $L_1=\max_{\bar{x}_s,\bar{u}_s\in\mathcal{S}_{\tau}^{\text{cur}}}\parallel g(\bar{x}_{s+1},\theta_\tau)-A_\tau g(\bar{x}_s,\theta_\tau)-B_\tau\bar{u}_s\parallel.$
    We denote the right-hand side of the above inequality~\eqref{eq:C12} as \( L_b \).
    For the last two terms in the error expansion~\eqref{eq:C5}, it follows from Assumption~3 that $\|x_k-\bar{x}_{k-1}\|\leq\mu_x$, and since $\bar{\epsilon}_k$ is defined on $\mathcal{S}_{\tau}^{\text{cur}}$, one can compute its upper bound by
    \begin{equation}\label{eq:C13}
        \parallel\bar{\epsilon}_{k-1}\parallel\leq E_{\text{recon}}=\max_{\bar{x}\in\mathcal{S}_{\tau}^{\text{cur}}}\parallel\bar{x}-C_\tau g(\bar{x},\theta_\tau)\parallel.
    \end{equation}
    Let $L_c:=\mu_x+E_{\text{recon}}$. Then the estimation error in~\eqref{eq:error_decomp} is upper bound by 
    \begin{equation}
        \|e_k\|\leq L_a+L_b+L_c.
    \end{equation}
    Following the proof of Theorem~1 in~\cite{haoDeepKoopmanLearning2024}, and based on Assumptions~\ref{assum:orthonormal} together with Lemma~\ref{lem:convergence}, we obtain
    \begin{equation}
        \begin{aligned}
            \lim_{n_h \to \infty} L_1&=\max_{\bar{x}_s,\bar{u}_s\in\mathcal{S}_{\tau}^{\text{cur}}}\parallel g(\bar{x}_{s+1},\theta_\tau)-A_\tau g(\bar{x}_s,\theta_\tau)-B_\tau\bar{u}_s\parallel\\
            &=0.
        \end{aligned}
    \end{equation}
    Finally, recall $L_a,L_b$ and $L_c$ defined above, we obtain
    \begin{equation}
        \lim_{n_h\rightarrow\infty}\sup \|e_k\| \leq (\|C_{\tau}A_{\tau}\|\mu_g+1)\mu_x+\|C_{\tau}B_{\tau}\|\mu_u+E_{\text{recon}},
    \end{equation}
    where $E_{\text{recon}}$ is the reconstruction error in~\eqref{eq:C13}.
\end{pf}
Using~\eqref{eq:C12} to bound the local error by the global accumulated error has an important advantage: 
if we include a loss term involving \( \|A_{\tau}\|_2 \) in the neural-network training loss so that \( \|A_{\tau}\|_2 < 1 \), then the upper bound of \( \|e_k\| \) is determined by the minimization performance of $E_{\text{recon}}$ and the constants \( \mu_x \) and \( \mu_u \) as \( k \to \infty \), rather than by letting \( n_h \to \infty \). 
A detailed proof of this conclusion can be found in the proof of Corollary~1 in~\cite{haoDeepKoopmanLearning2024}.

\section{Proof of Lemma~\ref{thm:RoA}}\label{apdx:E}
\begin{pf}
    We consider a candidate input sequence $U(k)=[u_{k,k}=K\hat{g}_{k,k},u_{k+1,k}=K\hat{g}_{k+1,k},\ldots, u_{k+H-1,k}=K\hat{g}_{k+H-1,k}]$ together with the corresponding state sequence $X=[\hat{g}_{k,k},\ldots,\hat{g}_{k+H,k}]$.
    From~\eqref{eq:P_matrix}, we obtain:
    \begin{equation}\label{eq:E1}
        \hat{g}_{k,k}^\top P \hat{g}_{k,k}-\hat{g}_{k+1,k}^\top P\hat{g}_{k+1,k}\geq \hat{g}_{k,k}^\top Q\hat{g}_{k,k}+\hat{g}_{k,k}^\top K^{\top}RK\hat{g}_{k,k}.
    \end{equation}
    Thus, it follows that
    \begin{equation}
        \gamma\geq\left\|\hat{g}_{k,k}\right\|_P^2>\left\|\hat{g}_{k+1,k}\right\|_P^2>\ldots>\left\|\hat{g}_{k+H,k}\right\|_P^2.
    \end{equation}
    Hence, the predicted states $\hat{g}_{k+j,k},j=0,\ldots,H$ lie in the set $\Xi=\{\hat{g}\in\mathbb{R}^n\mid \hat{g}^{\top}P\hat{g}\leq\gamma\}$, i.e. the set $\Xi$ is a positive invariant set.
    We select the Lyapunov candidate function $V_f(\hat{g})=\hat{g}^\top P\hat{g}$, which is a positive definite function (since $P$ is positive definite).
    From~\eqref{eq:E1}, we can obtain:
    \begin{equation}
        \begin{aligned}
                &V_f(\hat{g}_{k+j+1,k})-V_f(\hat{g}_{k+j,k})\\
                \leq&-\left\|\hat{g}_{k+j,k}\right\|_Q^2-\left\|K\hat{g}_{k+j,k}\right\|_R^2< 0.
        \end{aligned}
    \end{equation}
    Therefore, $\Xi$ is the region of attraction of the nominal system $\hat{g}_{k+1}=A_{\tau}\hat{g}_k+B_{\tau}u_k$ under the MPC controller~\eqref{eq:mpc}.
\end{pf}

\section{The proof of Theorem~\ref{thm:recursive_feasibility}}\label{apdx:F}
\begin{pf}
    First, we show that if the SDP problem~\eqref{eq:SDP} is feasible at time instant $k$, then it remains feasible at time instant $k+1$.
    The proof of this part follows the proof of Theorem~5 in~\cite{chen2024learning}.
    Assume that the SDP~\eqref{eq:SDP} is feasible at time instant \( k \), with the optimal solution denoted as \( P_k, K_k \).    
    The states sequence $X(k):=[{g}_{k,k},\ldots,\hat{g}_{k+H,k}]\in\Xi$. Then, the candidate input sequence at time instant \( k \) is constructed as $U(k)=[K_k{g}_{k,k},K_k\hat{g}_{k+1,k},\ldots,K_k\hat{g}_{k+H-1,k}]$.
    Since $\epsilon_k=g(f(x_k,K{g}_k))-(A_{\tau}+B_{\tau}K_k){g}_k$, we have
    \begin{equation}
        \begin{aligned}
            &{g}_k^{\top}P _k{g}_k-g(f(x_k,K{g}_k))^{\top}P_kg(f(x_k,K{g}_k))\\
            =&{g}_k^{\top}P_k{g}_k-g(f(x_k,K{g}_k))^{\top}P_kg(f(x_k,K{g}_k))\\
            &\pm [(A_{\tau}+B_{\tau}K_k){g}_k]^{\top}P_kg(f(x_k,K{g}_k))\\
            &\pm [(A_{\tau}+B_{\tau}K_k){g}_k]^{\top}P_k[(A_{\tau}+B_{\tau}K_k){g}_k]\\
            \geq&{g}_k^{\top}P _k{g}_k-[(A_{\tau}+B_{\tau}K_k){g}_k]^{\top}P_k[(A_{\tau}+B_{\tau}K_k){g}_k]\\
            &-\left\|[g(f(x_k,K{g}_k))+(A_{\tau}+B_{\tau}K_k){g}_k]^{\top}P_k\epsilon_k.\right\|
        \end{aligned}
    \end{equation}
    Thus, at time instant $k+1$, following Assumption~\ref{assum:gamma}, we have
    \begin{equation}
        \begin{aligned}
            &g_k^{\top}P_k g_k-g(f(x_k,K{g}_k))^{\top}P_kg(f(x_k,K{g}_k))\\
            \geq&g_k^{\top}P_k g_k-\|(A_{\tau}+B_{\tau}K_k){g}_k\|_{P_k}^2-\mu_k \left\|\epsilon_k\right\|\\
            \overset{(a)}{\geq}&\|g_k\|_Q^2+\|Kg_k\|_R^2-\mu_k \left\|\epsilon_k\right\|\geq-\mu_k \left\|\epsilon_k\right\|,\\
        \end{aligned}
    \end{equation}
    where $(a)$ follows from~\eqref{eq:P_matrix}.
    It yields that
    \begin{equation}
        \begin{aligned}
            g_{k+1}^{\top}P_kg_{k+1}-\mu_k\left\|\epsilon_k\right\|\leq g_k^{\top}P_kg_k\overset{(b)}{\leq} \gamma-\mu_k\left\|\epsilon_k\right\|,\\
        \end{aligned}
    \end{equation}
    where $(b)$ follows from Assumption~\ref{assum:gamma}.
    Then, we obtain that $g_{k+1}^{\top}P_kg_{k+1}\leq \gamma$, which implies that $P_k$ is a candidate solution to the SDP~\eqref{eq:SDP} at time instant $k+1$.
    The remainder of the proof follows that of Theorem~5 in~\cite{chen2024learning}.
\end{pf}
\vspace{-0.4cm}
\section{The proof of Theorem~\ref{thm:ISS}}\label{apdx:G}
\vspace{-0.4cm}
\begin{pf}
We consider the closed-loop system composed of the exact lifted high-dimensional dynamics and the MPC controller in~\eqref{eq:mpc}.
Next, we show that the optimal value function of the MPC problem serves as an ISS Lyapunov function.
Suppose that the optimal input sequence at time instant $k$ is denoted by $U^*(k) = [u_{k,k}^*, \ldots, u_{k+H-1,k}^*]$, with the corresponding state sequence given by $X(k) = [\hat{g}_{k,k}^*, \ldots, \hat{g}_{k+H,k}^*]$.
The recursive feasibility of the MPC problem is guaranteed by Theorem~\ref{thm:recursive_feasibility}.
At time instant $k+1$ we select a feasible candidate control input sequence $U(k+1) = [u_{k+1,k}^*,\ldots, u_{k+H-1,k}^*,K_k\hat{g}_{k+H,k}^*]$, which can be readily verified to be a feasible solution to the MPC problem.
{The optimal value function at time instant $k$ can be expressed as}
\vspace{-0.3cm}
\begin{equation}
    V^*(k,\hat{g}_{k}) = \sum_{i=0}^{H-1}(\|\hat{g}_{k+i,k}^*\|_Q^2+\|u_{k+i,k}^*\|_R^2)+\|\hat{g}_{k+H,k}^*\|_P^2.
\end{equation}
First, we need to show that $\|\hat{g}\|_Q^2\leq V^*(k,\hat{g}_k) \leq \|\hat{g}_k\|_P^2$.
From the definition of the value function, it is straightforward to see that $\|\hat{g}_k\|_Q^2 \le V^*(k,\hat{g}_k)$.
We construct a feasible control sequence at time instant $k$ as $U(k) = [K_k\hat{g}_{k,k}, \ldots, K_k\hat{g}_{k+1,k},\ldots, K_k\hat{g}_{k+H-1,k}]$.
The value function using this feasible solution is given by
\vspace{-0.3cm}
\begin{equation}
    \begin{aligned}
            &V^*(k,\hat{g}_k)\leq V(k,\hat{g}_k) \\
            =&\sum_{i=0}^{H-1}(\|\hat{g}_{k+i,k}\|_Q^2+\|K_k\hat{g}_{k+i,k}\|_R^2)+\|\hat{g}_{k+H,k}\|_P^2\\
            \overset{(a)}{\leq}&\sum_{i=0}^{H-1}(\|\hat{g}_{k+i,k}\|_P^2-\|\hat{g}_{k+i+1,k}\|_P^2)+\|\hat{g}_{k+H,k}\|_P^2\overset{(b)}{=}\|\hat{g}_k\|_P^2,\\
    \end{aligned}
\end{equation}
where (a) follows from~\eqref{eq:P_matrix} and (b) results from that the expanded terms cancel each other out.
Hence, we have $\|\hat{g}_k\|_Q^2 \leq V^*(k,\hat{g}_k) \leq \|\hat{g}_k\|_P^2$.
This indicates that $\beta_1(\|\hat{g}_k\|)\leq V^*(k,\hat{g}_k)\leq \beta_2(\|\hat{g}_k\|)$, where $\beta_1(s)=\lambda_{min}(Q)s^2$ and $\beta_2(s)=\lambda_{max}(P)s^2$ are class $\mathscr{K}_\infty$ functions.
Moreover, we have
\vspace{-0.3cm}
\begin{equation}\label{eq:D3}
    \sum_{i=0}^{H-1}\|\hat{g}_{k+i,k}\|_Q^2\leq V^*(k,\hat{g}_k)\leq\|\hat{g}_{k}\|_P^2.
\end{equation}
At time instant $k+1$, we have $\hat{g}_{k+1}=g(f(x_k,u_k^*,k))=A_{\tau}\hat{g}_{k}+B_{\tau}u_{k}^*+\epsilon_{k}$, where $\epsilon_{k}$ is the approximation error of the Koopman operator.
By applying the feasible solution $U(k+1) = [u_{k+1,k}^*,\ldots, u_{k+H-1,k}^*,K_k\hat{g}_{H,k}]$, the corresponding lifted state sequence at $k+1$ is given by
\vspace{-0.6cm}{\small
\begin{equation}
    \begin{aligned}
        \hat{g}_{k+1,k+1}&=A_{\tau}\hat{g}_{k}+B_{\tau}u_{k}^*+\epsilon_{k}=\hat{g}_{k+1,k}^*+\epsilon_{k}\\
        \hat{g}_{k+2,k+1}&=A_{\tau}\hat{g}_{k+1,k+1}+B_{\tau}u_{k+1,k}^*\\
        &= A_{\tau}\hat{g}_{k+1,k}+B_{\tau}u_{k+1,k}^*+A_{\tau}\epsilon_{k}=\hat{g}_{k+2,k}^*+A_{\tau}\epsilon_{k}\\
        &\vdots\\
        \hat{g}_{k+H+1,k+1}&=A_{\tau}\hat{g}_{k+H,k+1}+B_{\tau}K_k\hat{g}_{k+H,k}^*\\
        &= A_{\tau}\hat{g}_{k+H,k}^*+B_{\tau}K_k\hat{g}_{k+H,k}^*+A_{\tau}^{H}\epsilon_{k}\\
    \end{aligned}
\end{equation}}
The $V(k+1,\hat{g}_{k+1})$ associated with $U(k+1)$ is given by
\begin{equation}
    \begin{aligned}
        &V^*(k+1,\hat{g}_{k+1}) \leq V(k+1,\hat{g}_{k+1}) \\
        =&\sum_{i=0}^{H-1}\|\hat{g}_{k+1+i,k}^*+A_{\tau}^i \epsilon_{k}\|_Q^2+\sum_{i=0}^{H-2}\|u_{k+i+1,k}^*\|_R^2+\\
        &\|K_k\hat{g}_{k+H,k}^*\|_R^2+\|A_{\tau}\hat{g}_{k+H,k}^*+B_{\tau}K_k\hat{g}_{k+H,k}^*+A_{\tau}^{H}\epsilon_{k}\|_P^2\\
        \overset{(c)}{\leq}&\sum_{i=0}^{H-2}(\|\hat{g}_{k+1+i,k}^*\|_Q^2+\|u_{k+i+1,k}^*\|_R^2)+\|\hat{g}_{k+H,k}^*\|_Q^2\\
        &+\|K_k\hat{g}_{k+H,k}^*\|_R^2+\|(A_{\tau}+B_{\tau}K_k)\hat{g}_{k+H,k}^*\|_P^2\\
        &+\sum_{i=0}^{H-1}(\eta\|\hat{g}_{k+1+i,k}^*\|_Q^2+(1+1/\eta)\|A_{\tau}^i \epsilon_{k}\|_Q^2)+\|A_{\tau}^H\epsilon_{k}\|_P^2\\
        \overset{(d)}{=}&V^*(k,\hat{g}_k)-\|\hat{g}_{k,k}^*\|_Q^2-\|u_{k,k}^*\|_R^2-\|\hat{g}_{k+H,k}^*\|_P^2+\\
        &\|\hat{g}_{k+H,k}^*\|_Q^2+\|K_k\hat{g}_{k+H,k}^*\|_R^2+\|(A_{\tau}+B_{\tau}K_k)\hat{g}_{k+H,k}^*\|_P^2\\
        &+\sum_{i=0}^{H-1}(\eta\|\hat{g}_{k+1+i,k}^*\|_Q^2+(1+1/\eta)\|A_{\tau}^i \epsilon_{k}\|_Q^2)+\|A_{\tau}^H\epsilon_{k}\|_P^2\\
        \overset{(e)}{\leq}&V^*(k,\hat{g}_k)+\sigma(\|\epsilon_k\|)-\|\hat{g}_{k,k}^*\|_Q^2+\eta{\lambda_{max}(P)}\|\hat{g}_{k,k}^*\|^2\\
        \overset{(f)}{\leq}&V^*(k,\hat{g}_k)+\sigma(\|\epsilon_k\|)-(\lambda_{min}(Q)-\eta{\lambda_{max}(P)})\|\hat{g}_{k,k}^*\|^2,
    \end{aligned}
\end{equation}
where $(c)$ follows Young's inequality (i.e., $\|\hat{g}_{k}+\epsilon_{k}\|_Q^2\leq(1+\eta)\|\hat{g}_{k}\|_Q^2+(1+1/\eta)\|\epsilon_{k}\|_Q^2$), $(d)$ follows from $V^*(k,\hat{g}_k)=\sum_{i=0}^{H-1}(\|\hat{g}_{k+i,k}\|_Q^2+\|u_{k+i,k}\|_R^2)+\|\hat{g}_{k+H,k}\|_P^2$ and $(f)$ follows from $-\|\hat{g}_k\|_Q^2\leq-\lambda_{min}(Q)\|\hat{g}_k\|_Q^2$.
The above inequality for $(e)$ arises from three components: The first term follows directly from~\eqref{eq:P_matrix}; The second term is obtained, similar to~\cite{chen2024learning}, by aggregating all terms related to $\epsilon_k$ into $\sigma(\epsilon_k)$; The third term is given by $\sum_{i=0}^{H-1} \eta \lVert \hat{g}^*_{k+1+i,k} \rVert_Q^2\overset{\eqref{eq:D3}}{\leq} {\eta}\|\hat{g}_{k+1,k}\|_P^2\overset{\text{Lemma~\ref{thm:RoA}}}{\leq} {\eta}\|\hat{g}_{k}\|_P^2\leq \eta{\lambda_{max}(P)}\|\hat{g}_k\|^2$.
As long as $\eta$ in the Young's inequality relaxation is chosen sufficiently small such that $\lambda_{min}(Q)-\eta{\lambda_{max}(P)} > 0$, we obtain
\vspace{-0.2cm}
\begin{equation}
    V^*(k+1,\hat{g}_{k+1})-V^*(k,\hat{g}_k)\leq -\beta_3(\|\hat{g}_k\|)+\sigma(\|\epsilon_k\|).
\end{equation}
{Note that around the update instants of the lifted system matrices (i.e., when $A_\tau$ is updated to $A_{\tau+1}$), for the sake of proof simplicity, the error term $\epsilon_k$ is defined to include not only the Koopman model approximation error but also the discrepancy induced by model variation, namely $(A_{\tau+1}-A_\tau) g_{k+1} + (B_{\tau+1}-B_\tau) u_{k+1}$.
}After this discrepancy term is absorbed into $\epsilon_{k+1}$, the proof can be completed by following the same steps as above.
Finally, we have shown that the optimal value function $V^*(k,\hat{g}_k)$ is an ISS Lyapunov function; therefore, the closed-loop system is input-to-state stable.
\end{pf}

\bibliographystyle{unsrt}        
\bibliography{references}           

\end{document}